\documentclass[a4paper,11pt]{article}
\pdfoutput=1
\usepackage{jheppub}

\usepackage[T1]{fontenc}
\usepackage{amssymb}
\usepackage{amsmath}
\usepackage{graphicx}
\usepackage{subcaption}
\usepackage[colorlinks=true]{hyperref}
\usepackage{appendix}
\usepackage{amsbsy}

\def\beq{\begin{equation}}   
\def\eeq{\end{equation}}
\def\bea{\begin{eqnarray}}  
\def\eea{\end{eqnarray}} 

\def\eps{\epsilon}

\def\doubletilde#1{\widetilde{\vphantom{\raise 1.5pt \hbox{#1}}\smash{\kern -2pt\widetilde{#1}}}}

\def\NF{N_F}

\def\Re{\mbox{Re}}

\def\e{\epsilon}
\def\d{\hbox{d}}

\def\NF{N_F}

\def\e{\epsilon}
\def\d{\hbox{d}}

\newcommand{\code}[1]{\textsc{#1}}

\newcommand{\NFZ}{N_{F,\gamma}}

\preprint{{\raggedleft%
ZU-TH 54/22 
}}

\title{The parton-level structure of $e^+e^-$ to 2 jets at N$^3$LO}
\author{Petr Jakub\v{c}\'{i}k,}
\author{Matteo Marcoli and}
\author{Giovanni Stagnitto}

\affiliation{Physik-Institut, Universit\"at Z\"urich,
  Winterthurerstrasse 190, CH-8057 Z\"urich, Switzerland}
\emailAdd{petr.jakubcik@physik.uzh.ch}
\emailAdd{matteo.marcoli@physik.uzh.ch}
\emailAdd{giovanni.stagnitto@physik.uzh.ch}


\abstract{ We investigate the quantum chromodynamics (QCD) corrections to
  hadronic final states in electron-positron collisions at
  $\mathcal{O}(\alpha_s^3)$ in the strong coupling constant $\alpha_s$. Namely,
  we analytically compute the total cross section for this process by separately
  integrating the tree-level five-parton, the one-loop four-parton, the two-loop
  three-parton, and the three-loop two-parton matrix elements over the
  respective phase space. All the contributions to the calculation are treated
  in a common framework whereby phase space integrals are expressed as physical
  cuts of the four-loop two-point function. We check the cancellation of
  infrared poles at all colour levels and we reproduce the known result for the
  $R$-ratio at order $\alpha_s^3$. }

\begin{document} 
\maketitle
\flushbottom


\section{Introduction}

Hadronic final states at electron-positron colliders constitute the cleanest
environment for testing quantum chromodynamics (QCD).
Historically, measurements of the $R$-ratio, defined as the ratio of the cross
sections for hadron production and muon
production in $e^+e^-$ collisions mediated by photon exchange,
\begin{equation}\label{eq:Rratio}
  R = \frac{\sigma(e^+e^- \to \gamma^* \to \text{hadrons})}{\sigma(e^+e^- \to \gamma^* \to \mu^+\mu^-)} = N \sum_q e_q^2\, \Big( 1 + \mathcal{O}(\alpha_s) \Big)\,,
\end{equation}
with $N$ the number of colours and $e_q$ the electric charge of the quark,
supplied the first experimental evidence of the fractional charge of quarks and
for the existence of the colour quantum number.
From a theoretical point of view, this process provides the ideal playground for studying
the dynamics of strong interactions, given the colourless nature of the initial
state and the absence of QCD vertices at Born level.

The total cross section can be written as a perturbative expansion in terms of
the QCD strong coupling constant $\alpha_s$,
\begin{equation}\label{eqn:totalcross}
  \sigma(e^+e^- \to \gamma^* \to \text{hadrons}) = \sigma^{(0)} + \left(\frac{\alpha_s}{2\pi}\right)
  \sigma^{(1)} + \left(\frac{\alpha_s}{2\pi}\right)^2 \sigma^{(2)}
  + \left(\frac{\alpha_s}{2\pi}\right)^3 \sigma^{(3)} + \mathcal{O}(\alpha_s^4)\,,
\end{equation}
with $\sigma^{(0)}$ the Born-level or leading-order (LO) cross section, and
$\sigma^{(k)}$ denoting the N$^{k}$LO correction to the cross section.
The N$^{3}$LO ~\cite{Gorishnii:1990vf,Surguladze:1990tg} and the N$^{4}$LO
corrections~\cite{Baikov:2008jh,Baikov:2010je,Baikov:2011pkl,Baikov:2012er,Herzog:2017dtz}
are known.

Generally speaking, evaluating the terms in \eqref{eqn:totalcross} beyond Born
level requires the calculation of Feynman diagrams with additional real and
virtual particles. The emission of real particles causes singular behaviour of
the matrix element in the collinear and/or soft limit, while the exchange of
virtual particles leads to a divergent loop integration. The real and virtual
divergences can, however, be consistently treated separately after the
introduction of a regulator. The most common choice is dimensional
regularisation where phase space integrals as well as loop integrals are
performed in $d=4-2\e$ dimensions. This leads to the presence of explicit
$1/\e^n$ poles in the intermediate steps of the calculation. Nonetheless, the
KLN theorem~\cite{Kinoshita:1962ur,Lee:1964is} guarantees cancellation of
infrared (IR) poles for sufficiently inclusive observables -- such as the total
cross section -- and thus one can eventually take the limit $\e \to 0$.

In~\cite{Gehrmann-DeRidder:2004ttg} the infrared structure of the process $e^+e^- \to
2\,\rm{jets}$ was scrutinized up to NNLO.
Namely, the tree-level four-parton, the one-loop three-parton and the two-loop
two-parton matrix elements were integrated over the respective phase space and
summed to obtain the NNLO contribution to the cross section,
\begin{equation}\label{eqn:matelemsNNLO}
  \sigma^{(2)} = \int \d\Phi_4\,M_{4}^0
  + \int \d\Phi_3\,M_{3}^1
  + \int \d\Phi_2\,M_{2}^2\,,
\end{equation}
where $M_{n}^l$ denotes the $l$-loop squared matrix element for the decay of
a virtual photon into $n$ final state QCD particles. The phase space $\d \Phi_n$ is defined as
\begin{equation}
\d \Phi_n = \frac{\d^{d-1} p_1}{2E_1 (2\pi)^{d-1}}\; \ldots \; \frac{\d^{d-1}
  p_n}{2E_n (2\pi)^{d-1}}\; (2\pi)^{d} \; \delta^d (q - p_1 - \ldots - p_n) \,.
\end{equation}
We refer to each of the terms in~\eqref{eqn:matelemsNNLO} as a \textit{layer} of
the calculation. In an N$^{k}$LO calculation, each layer can feature infrared
poles up to $1/\e^{2k}$, i.e.\ up to $1/\e^4$ at NNLO and up to $1/\e^6$ at
N$^{3}$LO, etc.
The authors of~\cite{Gehrmann-DeRidder:2004ttg} identified IR-singular terms in
the three- and four-parton final states with the IR singularities of the
two-parton final state. They further observed the cancellation of the
contribution from the one-loop soft gluon current between three- and four-parton
final states.

The purpose of the present work is to extend the analysis
of~\cite{Gehrmann-DeRidder:2004ttg} to the layers of the
N$^{3}$LO cross section. We analogously integrate the tree-level five-parton, the
one-loop four-parton, the two-loop three-parton and the three-loop two-parton
matrix elements over their respective phase space:
\begin{equation}\label{eqn:matelems}
  \sigma^{(3)} = \int \d\Phi_5\,M_{5}^0
  + \int \d\Phi_4\,M_{4}^1
  + \int \d\Phi_3\,M_{3}^2
  + \int \d\Phi_2\,M_{2}^3\,.
\end{equation}

This work forms part of the joint effort of the community towards N$^{3}$LO
calculations~\cite{Anastasiou:2015vya,Mistlberger:2018etf,Currie:2018fgr,Cieri:2018oms,Mondini:2019gid,Duhr:2021vwj,
  Duhr:2020seh,Duhr:2020sdp,Chen:2022lwc,Chen:2022cgv,Chen:2021vtu,Chen:2021isd,Baglio:2022wzu,Chen:2022vzo}.
Several steps have been taken towards a complete description of the universal
behaviour of matrix elements in unresolved configurations in N$^3$LO
calculations. These scenarios include single-unresolved limits of two-loop
amplitudes, double-unresolved limits of one-loop amplitudes, and
triple-unresolved limits of tree-level amplitudes. Some unresolved limits can be
described with the iteration of single and double unresolved structures,
while others require novel computations, e.g.\ single collinear limits of
two-loop amplitudes~\cite{Badger:2004uk,Duhr:2014nda}; two-loop current for the
emission of a soft gluon~\cite{Duhr:2013msa,Li:2013lsa,Dixon:2019lnw}; triple collinear limits of one-loop amplitudes~\cite{Catani:2003vu,Czakon:2022fqi}; double soft emission at one loop~\cite{Catani:2021kcy,Zhu:2020ftr,Czakon:2022dwk}; quadruple collinear splitting functions~\cite{DelDuca:2019ggv,DelDuca:2020vst}; and triple soft
emission in tree-level amplitudes~\cite{Catani:2019nqv,DelDuca:2022noh,Catani:2022hkb}.
At the integrated level, the interplay of the different unresolved limits is not yet
understood. The cancellation pattern between the components
of~\eqref{eqn:matelems} could shed more light on the way the implicit infrared
divergent behaviour translates to the integrated level. 

We anticipate the importance of the presented results for the development of a
local subtraction scheme at N$^{3}$LO. The matrix elements of $\gamma^* \to
q\bar{q}$ and their integrated versions have been used in the context of the
antenna subtraction scheme for NNLO
calculations~\cite{Gehrmann-DeRidder:2005btv}. The matrix elements themselves
(with a proper normalisation) can serve as subtraction terms at the real level
to remove implicit singularities due to unresolved radiation between a
quark-antiquark pair of hard emitters. On the other hand, the integrals of the
matrix elements over the phase space have been used to remove the explicit
singularities present in virtual matrix elements. The results of the present
work can therefore provide necessary ingredients for a future formulation of the
antenna subtraction scheme at N$^{3}$LO.

The remainder of this paper is organised as follows. In Sect.~\ref{sec:method},
we briefly describe our methods. In Sect.~\ref{sec:results}, we present results
for the integrated matrix elements in order of final-state multiplicity. In
Sect.~\ref{sec:checks}, we combine the expressions to recover the known
$R$-ratio at $\mathcal{O}(\alpha_s^3)$ and we comment on the results. 
We conclude in Sect.~\ref{sec:conclusions} and elaborate on
possible future directions.


\begin{figure}[t]
    \centering
    \begin{subfigure}[b]{0.22\linewidth}
        \includegraphics[width=\linewidth]{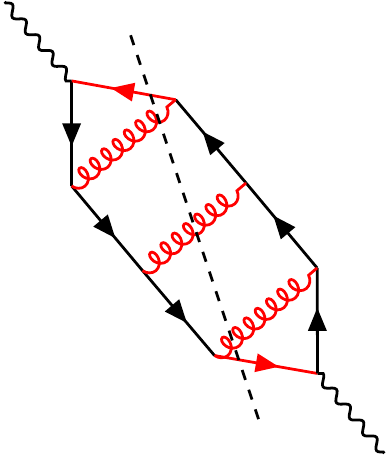}
        \caption{5 cut-propagators}
        \label{fig:a}
    \end{subfigure}
    \begin{subfigure}[b]{0.22\linewidth}
        \includegraphics[width=\linewidth]{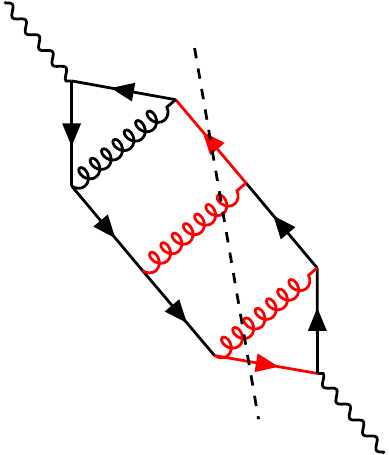}
        \caption{4 cut-propagators}
        \label{fig:b}
    \end{subfigure}
    \begin{subfigure}[b]{0.22\linewidth}
        \includegraphics[width=\linewidth]{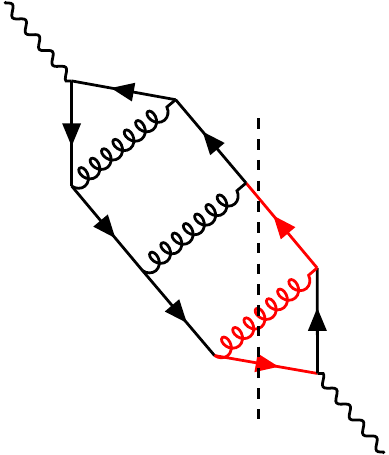}
        \caption{3 cut-propagators}
        \label{fig:c}
    \end{subfigure}
    \begin{subfigure}[b]{0.22\linewidth}
        \includegraphics[width=\linewidth]{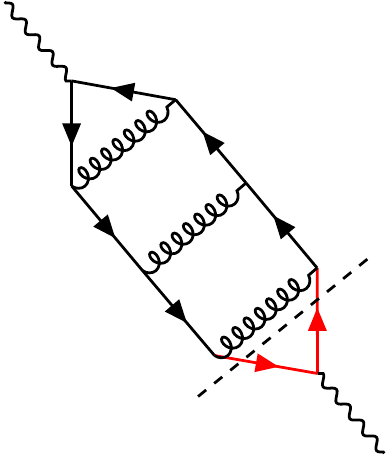}
        \caption{2 cut-propagators}
        \label{fig:d}
    \end{subfigure}
\caption{Example of a four-loop massless photon self-energy featuring three
  internal gluons with different possible choices of cut propagators (in red).
  According to the number of propagators on cut, the same diagram can represent
  the integral over the phase space of: \subref{fig:a} a tree-level squared
  amplitude $\gamma \to q g g g \bar{q}$, \subref{fig:b} the interference
  between a one-loop amplitude $\gamma \to q g g \bar{q}$ with its tree-level,
  \subref{fig:c} the interference between a two-loop amplitude $\gamma \to q g
  \bar{q}$ with its tree-level, \subref{fig:d} the interference between a
  three-loop amplitude $\gamma \to q \bar{q}$ with its tree-level.
}\label{fig:cuts}
\end{figure}

\section{Method}
\label{sec:method}

Phase-space integrals are related to loop integrals by the reverse unitarity
relation~\cite{Cutkosky:1960sp,Anastasiou:2002yz} which reformulates the mass-shell condition for $p$, a final-state massless momentum, as an inverse propagator on cut ({\em cut-propagator}),
\begin{equation}
  2\pi i\delta^+(p^2) = \frac{1}{p^2 - i0} - \frac{1}{p^2 + i0}\,.
\end{equation}
This way, all the integrals
present in the matrix elements in \eqref{eqn:matelems} can be expressed as cuts of the four-loop
photon self-energy and we can leverage techniques developed for the computation of loop integrals. According to the number of propagators on cut, a given self-energy diagram contributes to a layer in \eqref{eqn:matelems}. Namely, $n$-particle cuts represent the integration over the inclusive phase space  of a matrix element for the emission of $n$ real particles with $(5-n)$ loops. The relevant cuts of a particular diagram are depicted in Fig.~\ref{fig:cuts}.

At first, the four-loop diagrams with two external legs are generated with \code{QGRAF}~\cite{Nogueira:1991ex} using a model which includes the Standard Model QCD particles and couplings, as well as a set of fields for cut-propagators, which are also allowed to couple to regular particles. However, most naive arrangements of the cut-fields within a four-loop diagram are unphysical. In fact, a diagram only contributes to the physical integrated cross section if the cuts divide it into exactly two connected graphs, each attached to an external photon current. Moreover, each contribution in \eqref{eqn:matelems} requires a specific number of cuts and loops on either side of the cut. Finally, only cuts which fulfill momentum conservation and do not contain a self-energy insertion on cut-propagators are retained. In this way, we had to evaluate 1592, 3114, 2556 and 1764 diagrams for the integrations of the matrix elements $M_{2}^{3}$, $M_{3}^{2}$, $M_{4}^{1}$ and $M_{5}^{0}$, respectively. 

We subsequently insert the Feynman rules into the selected diagrams, compute the colour algebra, and evaluate the traces of gamma matrices. After imposing the on-shellness condition for cut propagators and representing each integral as a combination of scalar integrals, we can use \code{Reduze2}~\cite{vonManteuffel:2012np} to sort the integrals into the auxiliary topologies reported in Table~\ref{tab:auxtopo}.

\begin{table}
  \centering
  \begin{tabular}{cc|cc|cc}
    \multicolumn{2}{c}{F1} & \multicolumn{2}{c}{F2}  & \multicolumn{2}{c}{F3} \\
    \hline
    \hline
    $k_1$      & $k_2-k_3$ & $k_1$      & $k_2-k_3$       & $k_1$      & $k_4-k_2$          \\
    $k_2$      & $k_2-k_4$ & $k_2$      & $k_2-k_3$       & $k_2$      & $k_3-k_4$          \\   
    $k_3$      & $k_3-k_4$ & $k_3$      & $k_3-k_4$       & $k_4$      & $q-k_1$            \\
    $k_4$      & $q-k_1$   & $k_4$      & $q-k_1$         & $k_1-k_2$  & $q-k_2$            \\
    $k_1-k_2$  & $q-k_2$   & $k_1-k_2$  & $q-k_2$         & $k_1-k_3$  & $q-k_3$            \\
    $k_1-k_3$  & $q-k_3$   & $k_1-k_3$  & $q-k_4$         & $k_1-k_4$  & $q+k_2-k_3$        \\
    $k_1-k_4$  & $q-k_4$   & $k_1-k_4$  & $q+k_3-k_4$     & $k_2-k_3$  & $q+k_4-k_3-k_1$ 
  \end{tabular}
  \caption{The three auxiliary topologies which label every diagram present in the calculation. The knowledge of cut propagators is re-inserted only after the matching of diagrams onto the topologies F1, F2 and F3. One can subsequently infer which combinations of cuts are required in each topology to accommodate every cut diagram.}\label{tab:auxtopo}
\end{table}
  
Keeping track of which propagators are cut in any diagram, we can define for every auxiliary topology (F1, F2, F3) a set of \textit{cut-families} which cover all the integrals appearing in the matrix elements. Since we re-introduced the cuts only after matching onto the topologies, we are left with redundancy in the definition of the cut-families, which is taken care of internally by \code{Reduze2}'s sector relation finder before the reduction.

The integrals appearing in the matrix elements have up to eleven propagators in the denominator and a maximum of four scalar products in the numerator, in line with the calculation of the three-loop quark form factor in \cite{Gehrmann:2010ue}. For each layer, we reduce all integrals to a set of master integrals, finding $22$, $27$, $35$ and $31$ master integrals for the two-, \mbox{three-}, four-, and five-particle final state respectively. These master integral are in a one-to-one correspondence with those analytically computed in ~\cite{Gituliar:2018bcr,Magerya:2019cvz}. Namely, up to trivial relabeling of loop momenta, we can directly use their results.

\code{FORM}~\cite{Vermaseren:2000nd,Kuipers:2012rf} together with
\code{Mathematica} and \code{Python} scripts were used extensively throughout the calculation.

\section{Results}
\label{sec:results}

We illustrate the general structure of the different partonic contributions and
we provide explicit expressions only for the new $\mathcal{O}(\alpha_s^3)$
results. The notation of~\cite{Gehrmann-DeRidder:2004ttg} is adopted throughout this section for ease
of reference.

We present results for the integration of renormalised squared amplitudes.  The
renormalisation of ultraviolet divergences is performed in the
$\overline{{\rm MS}}$ scheme by means of the replacement
\begin{equation}\label{eq:alfaren}
\alpha_0\,\mu_0^{2\e}\,S_\e = \alpha_s\,\mu^{2\e}\left[
1- \frac{\beta_0}{\e}\left(\frac{\alpha_s}{2\pi}\right) 
+\left(\dfrac{\beta_0^2}{\e^2}-\dfrac{\beta_1}{2\e}\right)\left(\frac{\alpha_s}{2\pi}\right)^2
+{\cal O}(\alpha_s^3) \right]\,,
\end{equation}
with $\alpha_0$ the bare coupling and $\mu_0^2$ the mass parameter introduced in
dimensional regularisation to maintain a dimensionless coupling in the bare QCD
Lagrangian density. Additionally, $\alpha_s$ is the renormalised coupling
evaluated at the renormalisation scale $\mu^2$ and
\begin{eqnarray}
\beta_0 &=& \frac{11 C_A - 2 \NF}{6}\,,\label{eq:beta0}\\
\beta_1 &=& \frac{17 C_A^2 - 5 C_A \NF -3 C_F \NF}{6}\,,\label{eq:beta1}\\
S_\e &=& (4\pi)^\e e^{-\e\gamma}\,,\qquad \mbox{with Euler constant }
\gamma = 0.5772\ldots
\end{eqnarray}
Henceforth, we fix the renormalisation scale to be $\mu^2=q^2$, i.e.\
$\alpha_s\equiv \alpha_s(q^2)$.
Explicit relations between unrenormalised and renormalised amplitudes are
provided in Appendix~\ref{app:ren}.

\subsection{Two-parton final state}

The two-parton contribution to~\eqref{eqn:matelems} is given by the QCD loop
corrections to the $\gamma^{*}\to q\bar{q}$ process:
\begin{eqnarray}
|{\cal M}\rangle_{q\bar q} &=& \sqrt{4\pi\alpha}e_q  \Bigg[
|{\cal M}^{(0)}\rangle_{q\bar q} 
+ \left(\frac{\alpha_s}{2\pi}\right) |{\cal M}^{(1)}\rangle_{q\bar q} 
+ \left(\frac{\alpha_s}{2\pi}\right)^2 |{\cal M}^{(2)}\rangle_{q\bar q}
\nonumber \\ && \phantom{\sqrt{4\pi\alpha}e_q  \Bigg[}
+ \left(\frac{\alpha_s}{2\pi}\right)^3 |{\cal M}^{(3)}\rangle_{q\bar q} 
+ {\cal O}(\alpha_s^4) \Bigg] \,,
\label{eq:renorme}
\end{eqnarray}
with $\alpha$ the electromagnetic coupling constant, $e_q$ the quark charge,
and $|{\cal M}^{(i)}\rangle$ the $i$-loop contributions to the renormalised
amplitude.
The squared amplitude, summed over spins, colours and quark flavours
is given by:
\begin{eqnarray}
\hspace{-0.15cm}\langle{\cal M}|{\cal M}\rangle _{q\bar q} &=&  \sum |{\cal M}(\gamma^* \to q\bar q)|^2  \nonumber \\
&=& 4\pi\alpha \sum_q e_q^2 \Bigg[
\langle{\cal M}^{(0)}|{\cal M}^{(0)}\rangle_{q\bar q} + \left(\frac{\alpha_s}{2\pi}\right)
2\,\Re\big[\langle{\cal M}^{(1)}|{\cal M}^{(0)}\rangle_{q\bar q}\big] \nonumber \\
&& + \left(\frac{\alpha_s}{2\pi}\right)^2
\left(2\,\Re\big[\langle{\cal M}^{(2)}|{\cal M}^{(0)}\rangle_{q\bar q}\big]+\langle{\cal M}^{(1)}|{\cal M}^{(1)}\rangle_{q\bar q}\right)
\nonumber \\ 
&& + \left(\frac{\alpha_s}{2\pi}\right)^3
   \left(2\,\Re\big[\langle{\cal M}^{(3)}|{\cal M}^{(0)}\rangle_{q\bar q}\big]  +
         2\,\Re\big[\langle{\cal M}^{(2)}|{\cal M}^{(1)}\rangle_{q\bar q}\big]\right) 
 + {\cal O}(\alpha_s^4) \Bigg] .
\end{eqnarray}
We define
\begin{eqnarray}
  {\cal T}^{(2)}_{q\bar q} 
   &=&  \int \d \Phi_2 \,
            \langle{\cal M}^{(0)}|{\cal M}^{(0)}\rangle_{q\bar q},  \\
  {\cal T}^{(4)}_{q\bar q} 
   &=&  \int \d \Phi_2 \,
       2\,\Re\big[\langle{\cal M}^{(1)}|{\cal M}^{(0)}\rangle_{q\bar q}\big], \\
  {\cal T}^{(6)}_{q\bar q} &=& {\cal T}^{(6,\left[2\times 0\right])}_{q\bar q}
                               + {\cal T}^{(6,\left[1\times 1\right])}_{q\bar q}\,, \\  
  {\cal T}^{(6,\left[2\times 0\right])}_{q\bar q} 
	&=&  \int \d \Phi_2 \,
                  2\,\Re\big[\langle{\cal M}^{(2)}|{\cal M}^{(0)}\rangle_{q\bar q}\big],  \\
   {\cal T}^{(6,\left[1\times 1\right])}_{q\bar q} 
   &=&  \int \d \Phi_2 \,
       \langle{\cal M}^{(1)}|{\cal M}^{(1)}\rangle_{q\bar q},  \\
  {\cal T}^{(8)}_{q\bar q} &=& {\cal T}^{(8,\left[3\times 0\right])}_{q\bar q}
                               + {\cal T}^{(8,\left[2\times 1\right])}_{q\bar q}\,, \\    
  {\cal T}^{(8,\left[3\times 0\right])}_{q\bar q} 
   &=&  \int \d \Phi_2 \,
            2\,\Re\big[\langle{\cal M}^{(3)}|{\cal M}^{(0)}\rangle_{q\bar q}\big], \\
  {\cal T}^{(8,\left[2\times 1\right])}_{q\bar q} 
   &=&  \int \d \Phi_2 \,
            2\,\Re\big[\langle{\cal M}^{(2)}|{\cal M}^{(1)}\rangle_{q\bar q}\big].
\end{eqnarray}
The lowest order contribution is given by
\begin{equation}
  {\cal T}^{(2)}_{q\bar q} = 4 N (1-\e) q^2 P_2\,,
\end{equation}
with $P_2$ the volume of the two-particle phase
space,
\begin{equation}
P_2  = \int \d \Phi_2 =
2^{-3+2\e}\, \pi^{-1+\e}\, \frac{\Gamma(1-\e)}{\Gamma(2-2\e)}\,
(q^2)^{-\e} \,.
\end{equation}
Note that in contrast with~\cite{Gehrmann-DeRidder:2004ttg}, our definition of
${\cal T}^{(2)}_{q\bar q}$ includes a factor of $P_2$ because we always integrate over the full phase space.
The one- and two-loop contributions ${\cal T}^{(4)}_{q\bar q}$ and
${\cal T}^{(6)}_{q\bar q}$ are given in~(4.6) and~(4.7)
of~\cite{Gehrmann-DeRidder:2004ttg} and in Appendix~\ref{app:A}. The colour
decompositions of ${\cal T}^{(8,\left[3\times 0\right])}_{q\bar q} $ and ${\cal T}^{(8,\left[2\times 1\right])}_{q\bar q} $ reads
\begin{eqnarray}
  \label{eq:X23ansatzA}	
  {\cal T}^{(8,\left[3\times 0\right])}_{q\bar q} &=& \left( N - \frac{1}{N}\right) {\cal T}^{(2)}_{q\bar q}\, \Bigg[
  N^2 {\cal T}^{(8,\left[3\times 0\right])}_{q\bar q}\Big|_{N^2} + {\cal T}^{(8,\left[3\times 0\right])}_{q\bar q}\Big|_{N^0}
  + \frac{1}{N^2} {\cal T}^{(8,\left[3\times 0\right])}_{q\bar q}\Big|_{1/N^2}  
  \nonumber \\ &&                       
  + \NF N\, {\cal T}^{(8,\left[3\times 0\right])}_{q\bar q}\Big|_{\NF N}
  + \frac{\NF}{N}\,{\cal T}^{(8,\left[3\times 0\right])}_{q\bar q}\Big|_{\NF/N}
  + \NF^2 \,{\cal T}^{(8,\left[3\times 0\right])}_{q\bar q}\Big|_{\NF^2} 
  \nonumber \\ &&                       
  + \NFZ \left(N - \frac{4}{N}\right) \,{\cal T}^{(8,\left[3\times 0\right])}_{q\bar q}\Big|_{\NFZ}
  \Bigg]\,,
\end{eqnarray}
and
\begin{eqnarray}
	\label{eq:X23ansatzB}	
	{\cal T}^{(8,\left[2\times 1\right])}_{q\bar q} &=& \left( N - \frac{1}{N}\right) {\cal T}^{(2)}_{q\bar q}\, \Bigg[
	N^2 {\cal T}^{(8,\left[2\times 1\right])}_{q\bar q}\Big|_{N^2} + {\cal T}^{(8,\left[2\times 1\right])}_{q\bar q}\Big|_{N^0}
	+ \frac{1}{N^2} {\cal T}^{(8,\left[2\times 1\right])}_{q\bar q}\Big|_{1/N^2}
	\nonumber \\ &&
	+ \NF N\, {\cal T}^{(8,\left[2\times 1\right])}_{q\bar q}\Big|_{\NF N}                         
	+ \frac{\NF}{N}\,{\cal T}^{(8,\left[2\times 1\right])}_{q\bar q}\Big|_{\NF/N}
	\Bigg]\,,
\end{eqnarray}
where
\begin{equation}
  \NFZ = \frac{(\sum_q e_q)^2}{\sum_q e_q^2}\,.
\end{equation}
The last term in \eqref{eq:X23ansatzA} features a singlet contribution proportional to $(\sum_q e_q)^2$ which always comes with the expected quartic Casimir $d^{abc} d^{abc} = N-4/N$.
The terms in~\eqref{eq:X23ansatzA} up to finite order in $\e$ are given by:
\begin{flalign}
	{\cal T}^{(8,\left[3\times 0\right])}_{q\bar{q}}\Big|_{N^2} = &
	+\frac{1}{\e^6}\left(-\frac{1}{24}\right)
	+\frac{1}{\e^5}\left(-\frac{7}{8}\right)
	+\frac{1}{\e^4}\left(-\frac{5299}{1296}+\frac{17}{96}\pi^2\right)
	\nonumber &\\&
	+\frac{1}{\e^3}\left(-\frac{751}{243}+\frac{2185}{1296}\pi^2-\frac{7}{12}\zeta_{3}\right)
	\nonumber &\\&
	+\frac{1}{\e^2}\left(\frac{3371}{15552}+\frac{45551}{15552}\pi^2-\frac{1387}{432}\zeta_{3}-\frac{2299}{20736}\pi^4\right)
	\nonumber &\\&
	+\frac{1}{\e}\left(\frac{6297767}{279936}-\frac{79421}{46656}\pi^2-\frac{523}{108}\zeta_{3}-\frac{72347}{103680}\pi^4+\frac{151}{54}\pi^2\zeta_{3}-\frac{631}{60}\zeta_{5}\right)
	\nonumber &\\&
	+\frac{80373631}{1679616}-\frac{12363151}{559872}\pi^2+\frac{5521}{972}\zeta_{3}-\frac{506143}{622080}\pi^4+\frac{5477}{576}\pi^2\zeta_{3}
	\nonumber &\\&
	-\frac{2191}{80}\zeta_{5}+\frac{208037}{3732480}\pi^6-\frac{163}{144}\zeta_{3}^2 + \mathcal{O}(\e), &
\end{flalign}
\begin{flalign}
	{\cal T}^{(8,\left[3\times 0\right])}_{q\bar{q}}\Big|_{N^0} = &
	+\frac{1}{\e^6}\left(\frac{1}{12}\right)
	+\frac{1}{\e^5}\left(\frac{17}{16}\right)
	+\frac{1}{\e^4}\left(\frac{811}{288}-\frac{3}{8}\pi^2\right)
	+\frac{1}{\e^3}\left(\frac{6185}{864}-\frac{499}{192}\pi^2-\frac{11}{24}\zeta_{3}\right)
	\nonumber &\\&
	+\frac{1}{\e^2}\left(\frac{28075}{2592}-\frac{20135}{3456}\pi^2-\frac{95}{18}\zeta_{3}+\frac{1573}{5760}\pi^4\right)
	\nonumber &\\&
	+\frac{1}{\e}\left(-\frac{13385}{15552}-\frac{97123}{10368}\pi^2-\frac{29}{32}\zeta_{3}+\frac{240217}{207360}\pi^4+\frac{593}{288}\pi^2\zeta_{3}+\frac{33}{40}\zeta_{5}\right)
	\nonumber &\\&
	-\frac{9295397}{93312}+\frac{66571}{7776}\pi^2+\frac{3919}{96}\zeta_{3}+\frac{2757217}{1244160}\pi^4+\frac{47}{32}\pi^2\zeta_{3}
	\nonumber &\\&
	-\frac{923}{180}\zeta_{5}-\frac{35857}{1088640}\pi^6+\frac{35}{8}\zeta_{3}^2 + \mathcal{O}(\e), &
\end{flalign}
\begin{flalign}
	{\cal T}^{(8,\left[3\times 0\right])}_{q\bar{q}}\Big|_{1/N^2} = &
	+\frac{1}{\e^6}\left(-\frac{1}{24}\right)
	+\frac{1}{\e^5}\left(-\frac{3}{16}\right)
	+\frac{1}{\e^4}\left(-\frac{25}{32}+\frac{19}{96}\pi^2\right)
	\nonumber &\\&
	+\frac{1}{\e^3}\left(-\frac{83}{32}+\frac{53}{64}\pi^2+\frac{25}{24}\zeta_{3}\right)
	\nonumber &\\&
	+\frac{1}{\e^2}\left(-\frac{515}{64}+\frac{1273}{384}\pi^2+\frac{69}{16}\zeta_{3}-\frac{649}{3840}\pi^4\right)
	\nonumber &\\&
	+\frac{1}{\e}\left(-\frac{9073}{384}+\frac{4015}{384}\pi^2+\frac{2119}{96}\zeta_{3}-\frac{3833}{7680}\pi^4-\frac{1457}{288}\pi^2\zeta_{3}+\frac{161}{40}\zeta_{5}\right)
	\nonumber &\\&
	-\frac{53675}{768}+\frac{70429}{2304}\pi^2+\frac{2669}{32}\zeta_{3}-\frac{67177}{46080}\pi^4-\frac{3665}{192}\pi^2\zeta_{3}
	\nonumber &\\&
	+\frac{2119}{80}\zeta_{5}-\frac{19301}{1741824}\pi^6-\frac{913}{48}\zeta_{3}^2 + \mathcal{O}(\e), &
\end{flalign}
\begin{flalign}
	{\cal T}^{(8,\left[3\times 0\right])}_{q\bar{q}}\Big|_{\NF N} = &
	+\frac{1}{\e^5}\left(\frac{1}{8}\right)
	+\frac{1}{\e^4}\left(\frac{1283}{1296}\right)
	+\frac{1}{\e^3}\left(-\frac{253}{3888}-\frac{419}{2592}\pi^2\right)
	\nonumber &\\&
	+\frac{1}{\e^2}\left(-\frac{6461}{3888}-\frac{671}{15552}\pi^2+\frac{47}{216}\zeta_{3}\right)
	\nonumber &\\&
	+\frac{1}{\e}\left(-\frac{240167}{34992}+\frac{99107}{46656}\pi^2-\frac{521}{1296}\zeta_{3}+\frac{3941}{103680}\pi^4\right)
	\nonumber &\\&
	+\frac{573737}{209952}+\frac{387179}{69984}\pi^2-\frac{4549}{1296}\zeta_{3}-\frac{14527}{69120}\pi^4
	\nonumber &\\&
	+\frac{41}{288}\pi^2\zeta_{3}-\frac{205}{72}\zeta_{5} + \mathcal{O}(\e), &
\end{flalign}
\begin{flalign}
	{\cal T}^{(8,\left[3\times 0\right])}_{q\bar{q}}\Big|_{\NF/N} = &
	+\frac{1}{\e^5}\left(-\frac{1}{8}\right)
	+\frac{1}{\e^4}\left(-\frac{35}{144}\right)
	+\frac{1}{\e^3}\left(-\frac{139}{432}+\frac{17}{96}\pi^2\right)
	\nonumber &\\&
	+\frac{1}{\e^2}\left(\frac{775}{1296}-\frac{137}{1728}\pi^2+\frac{55}{72}\zeta_{3}\right)
	\nonumber &\\&
	+\frac{1}{\e}\left(\frac{24761}{3888}-\frac{9859}{5184}\pi^2-\frac{469}{432}\zeta_{3}-\frac{577}{20736}\pi^4\right)
	\nonumber &\\&
	+\frac{691883}{23328}-\frac{87853}{7776}\pi^2-\frac{21179}{1296}\zeta_{3}+\frac{125143}{622080}\pi^4
	\nonumber &\\&
	+\frac{85}{96}\pi^2\zeta_{3}
	+\frac{193}{72}\zeta_{5} + \mathcal{O}(\e), &
\end{flalign}
\begin{flalign}
	{\cal T}^{(8,\left[3\times 0\right])}_{q\bar{q}}\Big|_{\NF^2} = &
	+\frac{1}{\e^4}\left(-\frac{11}{162}\right)
	+\frac{1}{\e^3}\left(-\frac{1}{243}\right)
	+\frac{1}{\e^2}\left(\frac{23}{324}+\frac{\pi^2}{108}\right)
	\nonumber &\\&
	+\frac{1}{\e}\left(\frac{2417}{17496}-\frac{5}{324}\pi^2-\frac{\zeta_{3}}{81}\right)
	\nonumber &\\&
	-\frac{190931}{104976}+\frac{403}{972}\pi^2-\frac{52}{243}\zeta_{3}+\frac{43}{9720}\pi^4 + \mathcal{O}(\e), &
\end{flalign}
\begin{flalign}
	{\cal T}^{(8,\left[3\times 0\right])}_{q\bar{q}}\Big|_{\NFZ} = &
	+\frac{1}{2}+\frac{5}{24}\pi^2+\frac{7}{12}\zeta_{3}-\frac{\pi^4}{720}-\frac{10  \zeta_{5}}{3} + \mathcal{O}(\e), &
\end{flalign}
The terms in~\eqref{eq:X23ansatzA} up to finite order in $\e$ are given by:
\begin{flalign}
	{\cal T}^{(8,\left[2\times 1\right])}_{q\bar{q}}\Big|_{N^2} = &
	+\frac{1}{\e^6}\left(-\frac{1}{8}\right)
	+\frac{1}{\e^5}\left(-\frac{5}{4}\right)
	+\frac{1}{\e^4}\left(-\frac{259}{72}+\frac{7}{96}\pi^2\right)
	+\frac{1}{\e^3}\left(-\frac{8507}{864}+\frac{37}{96}\pi^2\right)
	\nonumber &\\&
	+\frac{1}{\e^2}\left(-\frac{99269}{5184}+\frac{517}{432}\pi^2+\frac{253}{144}\zeta_{3}-\frac{11}{768}\pi^4\right)
	\nonumber &\\&
	+\frac{1}{\e}\left(-\frac{655493}{31104}+\frac{30709}{10368}\pi^2-\frac{527}{144}\zeta_{3}-\frac{55}{1152}\pi^4+\frac{11}{48}\pi^2\zeta_{3}-\frac{33}{10}\zeta_{5}\right)
	\nonumber &\\&
	+\frac{8695267}{186624}+\frac{303889}{62208}\pi^2-\frac{33707}{864}\zeta_{3}-\frac{11953}{34560}\pi^4+\frac{2365}{1728}\pi^2\zeta_{3}
	\nonumber &\\&
	-\frac{3839}{240}\zeta_{5}-\frac{389}{193536}\pi^6+\frac{95}{16}\zeta_{3}^2 + \mathcal{O}(\e), &
\end{flalign}
\begin{flalign}
	{\cal T}^{(8,\left[2\times 1\right])}_{q\bar{q}}\Big|_{N^0} = &
	+\frac{1}{\e^6}\left(\frac{1}{4}\right)
	+\frac{1}{\e^5}\left(\frac{29}{16}\right)
	+\frac{1}{\e^4}\left(\frac{1711}{288}-\frac{\pi^2}{6}\right)
	+\frac{1}{\e^3}\left(\frac{15311}{864}-\frac{143}{192}\pi^2-\frac{13}{8}\zeta_{3}\right)
	\nonumber &\\&
	+\frac{1}{\e^2}\left(\frac{28153}{648}-\frac{8843}{3456}\pi^2-\frac{313}{36}\zeta_{3}+\frac{59}{5760}\pi^4\right)
	\nonumber &\\&
	+\frac{1}{\e}\left(\frac{708011}{7776}-\frac{72937}{10368}\pi^2-\frac{8657}{288}\zeta_{3}+\frac{29}{23040}\pi^4+\frac{37}{32}\pi^2\zeta_{3}+\frac{9}{40}\zeta_{5}\right)
	\nonumber &\\&
	+\frac{6795617}{46656}-\frac{122945}{7776}\pi^2-\frac{89467}{864}\zeta_{3}-\frac{1787}{138240}\pi^4+\frac{3367}{864}\pi^2\zeta_{3}
	\nonumber &\\&
	+\frac{197}{60}\zeta_{5}+\frac{6647}{241920}\pi^6+\frac{125}{8}\zeta_{3}^2 + \mathcal{O}(\e), &
\end{flalign}
\begin{flalign}
	{\cal T}^{(8,\left[2\times 1\right])}_{q\bar{q}}\Big|_{1/N^2} = &
	+\frac{1}{\e^6}\left(-\frac{1}{8}\right)
	+\frac{1}{\e^5}\left(-\frac{9}{16}\right)
	+\frac{1}{\e^4}\left(-\frac{75}{32}+\frac{3}{32}\pi^2\right)
	\nonumber &\\&
	+\frac{1}{\e^3}\left(-\frac{63}{8}+\frac{23}{64}\pi^2+\frac{13}{8}\zeta_{3}\right)
	\nonumber &\\&
	+\frac{1}{\e^2}\left(-\frac{1555}{64}+\frac{523}{384}\pi^2+\frac{111}{16}\zeta_{3}+\frac{47}{11520}\pi^4\right)
	\nonumber &\\&
	+\frac{1}{\e}\left(-\frac{8957}{128}+\frac{391}{96}\pi^2+\frac{1079}{32}\zeta_{3}+\frac{119}{2560}\pi^4-\frac{133}{96}\pi^2\zeta_{3}+\frac{123}{40}\zeta_{5}\right)
	\nonumber &\\&
	-\frac{49215}{256}+\frac{2797}{256}\pi^2+\frac{2281}{16}\zeta_{3}+\frac{1837}{5120}\pi^4-\frac{337}{64}\pi^2\zeta_{3}
	\nonumber &\\&
	+\frac{1017}{80}\zeta_{5}-\frac{24643}{967680}\pi^6-\frac{345}{16}\zeta_{3}^2 + \mathcal{O}(\e), &
\end{flalign}
\begin{flalign}
	{\cal T}^{(8,\left[2\times 1\right])}_{q\bar{q}}\Big|_{\NF N} = &
	+\frac{1}{\e^5}\left(\frac{1}{8}\right)
	+\frac{1}{\e^4}\left(\frac{35}{144}\right)
	+\frac{1}{\e^3}\left(\frac{187}{432}-\frac{\pi^2}{96}\right)
	\nonumber &\\&
	+\frac{1}{\e^2}\left(-\frac{751}{1296}+\frac{41}{1728}\pi^2-\frac{23}{72}\zeta_{3}\right)
	\nonumber &\\&
	+\frac{1}{\e}\left(-\frac{58381}{7776}+\frac{1279}{5184}\pi^2+\frac{71}{144}\zeta_{3}-\frac{7}{2304}\pi^4\right)
	\nonumber &\\&
	-\frac{1801117}{46656}+\frac{21899}{15552}\pi^2+\frac{2617}{432}\zeta_{3}
	\nonumber &\\&
	+\frac{833}{69120}\pi^4-\frac{53}{864}\pi^2\zeta_{3}+\frac{5}{24}\zeta_{5} + \mathcal{O}(\e), &
\end{flalign}
\begin{flalign}
	{\cal T}^{(8,\left[2\times 1\right])}_{q\bar{q}}\Big|_{\NF/N} = &
	+\frac{1}{\e^5}\left(-\frac{1}{8}\right)
	+\frac{1}{\e^4}\left(-\frac{35}{144}\right)
	+\frac{1}{\e^3}\left(-\frac{187}{432}+\frac{\pi^2}{96}\right)
	\nonumber &\\&
	+\frac{1}{\e^2}\left(\frac{751}{1296}-\frac{41}{1728}\pi^2+\frac{23}{72}\zeta_{3}\right)
	\nonumber &\\&
	+\frac{1}{\e}\left(\frac{58381}{7776}-\frac{1279}{5184}\pi^2-\frac{71}{144}\zeta_{3}+\frac{7}{2304}\pi^4\right)
	\nonumber &\\&
	+\frac{1801117}{46656}-\frac{21899}{15552}\pi^2-\frac{2617}{432}\zeta_{3}
	\nonumber &\\&
	-\frac{833}{69120}\pi^4+\frac{53}{864}\pi^2\zeta_{3}-\frac{5}{24}\zeta_{5} + \mathcal{O}(\e), &
\end{flalign}
The presented expressions are related to the quark form factors up to three loops~\cite{Lee:2010cga,Baikov:2009bg,Gehrmann:2010ue}. In particular
\begin{eqnarray}
  \label{eq:X23int30}
  {\cal T}^{(8,\left[3\times 0\right])}_{q\bar{q}} &=& 
	\frac{1}{8} {\cal T}^{(2)}_{q\bar q} 
	\Bigg\{ \mathcal{F}^q_3\,\Re\left[\Delta(q^2)^3\right]
  - \frac{4\beta_0}{\e} \mathcal{F}^q_2\,\Re\left[\Delta(q^2)^2\right]
  \nonumber \\ && \phantom{\frac{1}{8} {\cal T}^{(2)}_{q\bar q} \Bigg\{}
       - \left(\frac{2\beta_1}{\e}-\frac{4 \beta_0^2}{\e^2}\right)
	\mathcal{F}^q_1\,\Re\left[\Delta(q^2)\right]
	\Bigg\}
\end{eqnarray}
and
\begin{equation}
  \label{eq:X23int12}
  {\cal T}^{(8,\left[2\times 1\right])}_{q\bar{q}}
	= \frac{1}{8} {\cal T}^{(2)}_{q\bar q} 
	\left\{
	\mathcal{F}^q_1 \mathcal{F}^q_2\,\Re\left[\Delta(q^2)\right]
	- \frac{2\beta_0}{\e} (\mathcal{F}^q_1)^2
	\right\}\,,
\end{equation}
where $\mathcal{F}^q_1$, $\mathcal{F}^q_2$ and $\mathcal{F}^q_3$ are
given by~(2.23),~(2.27) and~(5.4) of~\cite{Gehrmann:2010ue}. The factor
\begin{equation}
	\Delta(q^2) = (-1-i0)^{-\e}
\end{equation}
appears in time-like kinematics. The factor
1/8 in~\eqref{eq:X23int30} and~\eqref{eq:X23int12} compensates for the fact that
$\mathcal{F}^q_1$, $\mathcal{F}^q_2$ and $\mathcal{F}^q_3$ refer to an expansion
in $\alpha_s/(4\pi)$. Equations~\eqref{eq:X23int30} and~\eqref{eq:X23int12} are indeed satisfied after the insertion of the expressions from~\cite{Gehrmann:2010ue}, which is a strong check of our results.

\subsection{Three-parton final states}

The three-parton configuration receives contributions from the two
processes $\gamma^{*}\to~q\bar{q}g$ and $\gamma^{*}\to~ggg$. The three-gluon
final state only arises at order ${\cal O} (\alpha_{s}^3)$ as the interference
of two one-loop diagrams where three gluons are emitted from a fermionic
loop attached to the photon. The amplitudes are
\begin{eqnarray}
	|{\cal M}\rangle_{q\bar q g} = \sqrt{4\pi\alpha}e_q\, \sqrt{4\pi\alpha_s}\,
	\Bigg[&&|{\cal M}^{(0)}\rangle_{q\bar q g} 
	+ \left(\frac{\alpha_s}{2\pi}\right) |{\cal M}^{(1)}\rangle_{q\bar q g} \nonumber \\
	&&+ \left(\frac{\alpha_s}{2\pi}\right)^2 |{\cal M}^{(2)}\rangle_{q\bar q g}    
	+ {\cal O}(\alpha_s^3) \Bigg] 
\end{eqnarray}
and
\begin{equation}
	|{\cal M}\rangle_{g g g} = \sqrt{4\pi\alpha}e_q\, (4\pi\alpha_s)^{3/2}\,
	\left[|{\cal M}^{(1)}\rangle_{g g g} + {\cal O}(\alpha_s) \right] \,.
\end{equation}
The squared amplitudes summed over spins, colours and quark flavours, are
given by
\begin{eqnarray}
	\langle{\cal M}|{\cal M}\rangle_{q\bar q g} &=& \sum |{\cal M}(\gamma^* \to q\bar q g)|^2 \nonumber \\
	&=& 4\pi\alpha\sum_q e_q^2\,8\pi^2\Bigg[
      \left(\frac{\alpha_s}{2\pi}\right)\langle{\cal M}^{(0)}|{\cal M}^{(0)}\rangle_{q\bar q g}
  + \left(\frac{\alpha_s}{2\pi}\right)^2 2\,\Re\big[\langle{\cal M}^{(0)}|{\cal M}^{(1)}\rangle_{q\bar q g}\big]
  \nonumber \\  && + \left(\frac{\alpha_s}{2\pi}\right)^3
	 \left(2\,\Re\big[\langle{\cal M}^{(2)}|{\cal M}^{(0)}\rangle_{q\bar q g}\big]
	+ \langle{\cal M}^{(1)}|{\cal M}^{(1)}\rangle_{q\bar q g} \right)
	+ {\cal O}(\alpha_s^4) \Bigg] \,
\end{eqnarray}
and
\begin{eqnarray}
\langle{\cal M}|{\cal M}\rangle_{ggg} &=& \sum |{\cal M}(\gamma^* \to ggg)|^2 \nonumber \\
 &=& 4\pi\alpha\sum_q e_q^2\,512\pi^6\Bigg[\left(\frac{\alpha_s}{2\pi}\right)^3\langle{\cal M}^{(1)}|{\cal M}^{(1)}\rangle_{ggg}
+ {\cal O}(\alpha_s^4) \Bigg] \,.
\end{eqnarray}
The integrated matrix elements squared are defined as
\begin{eqnarray}
	{\cal T}^{(4)}_{q\bar q g} 
	 &=& 8\pi^2 \int \d \Phi_3 \,
	\langle{\cal M}^{(0)}|{\cal M}^{(0)}\rangle_{q\bar q g}\,, \\
	{\cal T}^{(6)}_{q\bar q g} 
	 &=& 8\pi^2 \int \d \Phi_3 \,
             2\,\Re\big[\langle{\cal M}^{(1)}|{\cal M}^{(0)}\rangle_{q\bar q g}\big]\,, \\
           {\cal T}^{(8)}_{q\bar q g} &=&
          {\cal T}^{(8,\left[2\times 0\right])}_{q\bar q g} +                                                         
           {\cal T}^{(8,\left[1\times 1\right])}_{q\bar q g}\,,\\
	{\cal T}^{(8,\left[2\times 0\right])}_{q\bar q g} 
	 &=& 8\pi^2 \int \d \Phi_3 \,
             2\,\Re\big[\langle{\cal M}^{(2)}|{\cal M}^{(0)}\rangle_{q\bar q g}\big]\,, \\
	{\cal T}^{(8,\left[1\times 1\right])}_{q\bar q g} 
	&=& 8\pi^2 \int \d \Phi_3 \,
	\langle{\cal M}^{(1)}|{\cal M}^{(1)}\rangle_{q\bar q g}\,, \\
	{\cal T}^{(8)}_{ggg} 
	 &=& 512\pi^6 \int \d \Phi_3 \,
	\langle{\cal M}^{(1)}|{\cal M}^{(1)}\rangle_{ggg}\,. 
\end{eqnarray}
Expressions for ${\cal T}^{(4)}_{q\bar q g}$ and ${\cal T}^{(6)}_{q\bar q g}$
are given in~(4.28) and~(4.33) of~\cite{Gehrmann-DeRidder:2004ttg} and in Appendix~\ref{app:A}. For ${\cal T}^{(8,\left[2\times 0\right])}_{q \bar q g}$ and ${\cal T}^{(8,\left[1\times 1\right])}_{q \bar q g}$ we refer to the following decompositions:
\begin{eqnarray}
	\label{eq:X32ansatz1A}	
	{\cal T}^{(8,\left[2\times 0\right])}_{q \bar q g} &=& \left( N - \frac{1}{N}\right) {\cal T}^{(2)}_{q\bar q}\, \Bigg[
	N^2 {\cal T}^{(8,\left[2\times 0\right])}_{q\bar{q}g}\Big|_{N^2} + {\cal T}^{(8,\left[2\times 0\right])}_{q \bar q g}\Big|_{N^0}
	+ \frac{1}{N^2} {\cal T}^{(8,\left[2\times 0\right])}_{q \bar q g}\Big|_{1/N^2}       
	\nonumber \\ &&                       
	+ \NF N\, {\cal T}^{(8,\left[2\times 0\right])}_{q \bar q g}\Big|_{\NF N} 
	+ \frac{\NF}{N}\,{\cal T}^{(8,\left[2\times 0\right])}_{q \bar q g}\Big|_{\NF/N}
	+ \NF^2 \,{\cal T}^{(8,\left[2\times 0\right])}_{q \bar q g}\Big|_{\NF^2}       
	\nonumber \\ &&                       
	+ \NFZ \left(N - \frac{4}{N}\right) \,{\cal T}^{(8,\left[2\times 0\right])}_{q \bar q g}\Big|_{\NFZ}
	\Bigg]\,,
\end{eqnarray}
\begin{eqnarray}
	\label{eq:X32ansatz1B}	
	{\cal T}^{(8,\left[1\times 1\right])}_{q \bar q g} &=& \left( N - \frac{1}{N}\right) {\cal T}^{(2)}_{q\bar q}\, \Bigg[
	N^2 {\cal T}^{(8,\left[1\times 1\right])}_{q\bar{q}g}\Big|_{N^2} + {\cal T}^{(8,\left[1\times 1\right])}_{q \bar q g}\Big|_{N^0}
	+ \frac{1}{N^2} {\cal T}^{(8,\left[1\times 1\right])}_{q \bar q g}\Big|_{1/N^2}      
	\nonumber \\ &&                       
	+ \NF N\, {\cal T}^{(8,\left[1\times 1\right])}_{q \bar q g}\Big|_{\NF N}  
	+ \frac{\NF}{N}\,{\cal T}^{(8,\left[1\times 1\right])}_{q \bar q g}\Big|_{\NF/N}
	+ \NF^2 \,{\cal T}^{(8,\left[1\times 1\right])}_{q \bar q g}\Big|_{\NF^2}
	\Bigg]\,.
\end{eqnarray}
For ${\cal T}^{(8,\left[2\times 0\right])}_{q \bar q g}$, explicit expressions for the different
contributions appearing in~\eqref{eq:X32ansatz1A} are:
\begin{flalign}
	{\cal T}^{(8,\left[2\times 0\right])}_{q\bar{q}g}\Big|_{N^2} = &
	+\frac{1}{\e^6}\left(\frac{29}{72}\right)
	+\frac{1}{\e^5}\left(\frac{128}{27}\right)
	+\frac{1}{\e^4}\left(\frac{23689}{1296}-\frac{343}{288}\pi^2\right)
	\nonumber &\\&
	+\frac{1}{\e^3}\left(\frac{497425}{7776}-\frac{44041}{5184}\pi^2-\frac{557}{72}\zeta_{3}\right)
	\nonumber &\\&
	+\frac{1}{\e^2}\left(\frac{2820559}{11664}-\frac{15847}{486}\pi^2-\frac{10645}{144}\zeta_{3}+\frac{99149}{103680}\pi^4\right)
	\nonumber &\\&
	+\frac{1}{\e}\Bigg(\frac{265414681}{279936}-\frac{5932087}{46656}\pi^2-\frac{61387}{216}\zeta_{3}
	\nonumber &\\&
	\phantom{+\frac{1}{\e}\bigg(\;\;}+\frac{960277}{207360}\pi^4+\frac{20885}{864}\pi^2\zeta_{3}-\frac{31687}{360}\zeta_{5}\Bigg)
	\nonumber &\\&
	+\frac{206830619}{52488}-\frac{143664697}{279936}\pi^2-\frac{186071}{162}\zeta_{3}+\frac{750137}{38880}\pi^4
	\nonumber &\\&
	+\frac{254009}{1728}\pi^2\zeta_{3}-\frac{178661}{240}\zeta_{5}-\frac{5842331}{26127360}\pi^6+\frac{14893}{144}\zeta_{3}^2 + \mathcal{O}(\e), &
\end{flalign}
\begin{flalign}
	{\cal T}^{(8,\left[2\times 0\right])}_{q\bar{q}g}\Big|_{N^0} = &
	+\frac{1}{\e^6}\left(-\frac{5}{8}\right)
	+\frac{1}{\e^5}\left(-\frac{245}{48}\right)
	+\frac{1}{\e^4}\left(-\frac{667}{36}+\frac{547}{288}\pi^2\right)
	\nonumber &\\&
	+\frac{1}{\e^3}\left(-\frac{30959}{432}+\frac{18439}{1728}\pi^2+\frac{135}{8}\zeta_{3}\right)
	\nonumber &\\&
	+\frac{1}{\e^2}\left(-\frac{1384741}{5184}+\frac{107647}{2592}\pi^2+\frac{14849}{144}\zeta_{3}-\frac{17101}{11520}\pi^4\right)
	\nonumber &\\&
	+\frac{1}{\e}\Bigg(-\frac{32266075}{31104}+\frac{5124899}{31104}\pi^2+\frac{173377}{432}\zeta_{3}
	\nonumber &\\&
	\phantom{+\frac{1}{\e}\bigg(\;\;}-\frac{1309403}{207360}\pi^4-\frac{16825}{288}\pi^2\zeta_{3}+\frac{5579}{24}\zeta_{5}\Bigg)
	\nonumber &\\&
	-\frac{795917293}{186624}+\frac{122157485}{186624}\pi^2+\frac{290617}{162}\zeta_{3}-\frac{4177081}{155520}\pi^4
	\nonumber &\\&
	-\frac{151987}{576}\pi^2\zeta_{3}+\frac{275473}{240}\zeta_{5}+\frac{2075989}{8709120}\pi^6-\frac{17695}{48}\zeta_{3}^2 + \mathcal{O}(\e), &
\end{flalign}
\begin{flalign}
	{\cal T}^{(8,\left[2\times 0\right])}_{q\bar{q}g}\Big|_{1/N^2} = &
	+\frac{1}{\e^6}\left(\frac{1}{4}\right)
	+\frac{1}{\e^5}\left(\frac{9}{8}\right)
	+\frac{1}{\e^4}\left(\frac{83}{16}-\frac{37}{48}\pi^2\right)
	\nonumber &\\&
	+\frac{1}{\e^3}\left(\frac{83}{4}-\frac{103}{32}\pi^2-\frac{103}{12}\zeta_{3}\right)
	\nonumber &\\&
	+\frac{1}{\e^2}\left(\frac{1301}{16}-\frac{933}{64}\pi^2-\frac{265}{8}\zeta_{3}+\frac{3289}{5760}\pi^4\right)
	\nonumber &\\&
	+\frac{1}{\e}\left(\frac{122941}{384}-\frac{2093}{36}\pi^2-\frac{2597}{16}\zeta_{3}+\frac{4877}{2304}\pi^4+\frac{4211}{144}\pi^2\zeta_{3}-\frac{6449}{60}\zeta_{5}\right)
	\nonumber &\\&
	+\frac{83007}{64}-\frac{66487}{288}\pi^2-\frac{33883}{48}\zeta_{3}+\frac{68713}{7680}\pi^4+\frac{3399}{32}\pi^2\zeta_{3}
	\nonumber &\\&
	-\frac{15757}{40}\zeta_{5}+\frac{14593}{622080}\pi^6+\frac{4943}{24}\zeta_{3}^2 + \mathcal{O}(\e), &
\end{flalign}
\begin{flalign}
	{\cal T}^{(8,\left[2\times 0\right])}_{q\bar{q}g}\Big|_{\NF N} = &
	+\frac{1}{\e^5}\left(-\frac{115}{216}\right)
	+\frac{1}{\e^4}\left(-\frac{1439}{648}\right)
	+\frac{1}{\e^3}\left(-\frac{20489}{3888}+\frac{739}{1296}\pi^2\right)
	\nonumber &\\&
	+\frac{1}{\e^2}\left(-\frac{189505}{11664}+\frac{5747}{3888}\pi^2+\frac{64}{9}\zeta_{3}\right)
	\nonumber &\\&
	+\frac{1}{\e}\left(-\frac{789553}{17496}+\frac{55595}{23328}\pi^2+\frac{1195}{54}\zeta_{3}-\frac{193}{3240}\pi^4\right)
	\nonumber &\\&
	-\frac{49737247}{419904}+\frac{68021}{139968}\pi^2+\frac{31019}{648}\zeta_{3}+\frac{26777}{77760}\pi^4-\frac{1501}{216}\pi^2\zeta_{3}
	\nonumber &\\&
	+\frac{190}{3}\zeta_{5} + \mathcal{O}(\e), &
\end{flalign}
\begin{flalign}
	{\cal T}^{(8,\left[2\times 0\right])}_{q\bar{q}g}\Big|_{\NF/N} = &
	+\frac{1}{\e^5}\left(\frac{5}{12}\right)
	+\frac{1}{\e^4}\left(\frac{71}{72}\right)
	+\frac{1}{\e^3}\left(\frac{649}{216}-\frac{193}{432}\pi^2\right)
	\nonumber &\\&
	+\frac{1}{\e^2}\left(\frac{19691}{2592}-\frac{1489}{2592}\pi^2-\frac{227}{36}\zeta_{3}\right)
	\nonumber &\\&
	+\frac{1}{\e}\left(\frac{263363}{15552}-\frac{7315}{15552}\pi^2-\frac{2051}{216}\zeta_{3}+\frac{1661}{51840}\pi^4\right)
	\nonumber &\\&
	+\frac{3046295}{93312}+\frac{480743}{93312}\pi^2-\frac{27095}{1296}\zeta_{3}-\frac{60871}{311040}\pi^4
	\nonumber &\\&
	+\frac{775}{144}\pi^2\zeta_{3}-\frac{3019}{60}\zeta_{5} + \mathcal{O}(\e), &
\end{flalign}
\begin{flalign}
	{\cal T}^{(8,\left[2\times 0\right])}_{q\bar{q}g}\Big|_{\NF^2} = &
	+\frac{1}{\e^4}\left(\frac{1}{12}\right)
	+\frac{1}{\e^3}\left(\frac{1}{8}\right)
	+\frac{1}{\e^2}\left(\frac{19}{48}-\frac{7}{144}\pi^2\right)
	\nonumber &\\&
	+\frac{1}{\e}\left(\frac{109}{96}-\frac{7}{96}\pi^2-\frac{25}{36}\zeta_{3}\right)
	\nonumber &\\&
	+\frac{213}{64}-\frac{133}{576}\pi^2-\frac{25}{24}\zeta_{3}-\frac{71}{17280}\pi^4 + \mathcal{O}(\e), &
\end{flalign}
\begin{flalign}
	{\cal T}^{(8,\left[2\times 0\right])}_{q\bar{q}g}\Big|_{\NFZ} = &
	-\frac{15}{4}-\frac{2}{9}\pi^2+\frac{\zeta_{3}}{6}+\frac{\pi^4}{360}-\frac{\pi^2\zeta_{3}}{3}+\frac{25  \zeta_{5}}{3} + \mathcal{O}(\e). &
\end{flalign}
The results for ${\cal T}^{(8,\left[1\times 1\right])}_{q \bar q g}$ are:
\begin{flalign}
	{\cal T}^{(8,\left[1\times 1\right])}_{q\bar{q}g}\Big|_{N^2} = &
	+\frac{1}{\e^6}\left(\frac{29}{72}\right)
	+\frac{1}{\e^5}\left(\frac{71}{24}\right)
	+\frac{1}{\e^4}\left(\frac{1879}{144}-\frac{373}{864}\pi^2\right)
	\nonumber &\\&
	+\frac{1}{\e^3}\left(\frac{851}{16}-\frac{659}{192}\pi^2-\frac{685}{72}\zeta_{3}\right)
	\nonumber &\\&
	+\frac{1}{\e^2}\left(\frac{63401}{288}-\frac{6403}{432}\pi^2-\frac{8585}{144}\zeta_{3}-\frac{2737}{34560}\pi^4\right)
	\nonumber &\\&
	+\frac{1}{\e}\left(\frac{135041}{144}-\frac{74453}{1152}\pi^2-\frac{30355}{108}\zeta_{3}+\frac{809}{23040}\pi^4+\frac{9749}{864}\pi^2\zeta_{3}-\frac{12349}{120}\zeta_{5}\right)
	\nonumber &\\&
	+\frac{530995}{128}-\frac{243419}{864}\pi^2-\frac{124439}{96}\zeta_{3}-\frac{48167}{51840}\pi^4+\frac{42359}{576}\pi^2\zeta_{3}
	\nonumber &\\&
	-\frac{142601}{240}\zeta_{5}-\frac{94961}{967680}\pi^6+\frac{18773}{144}\zeta_{3}^2 + \mathcal{O}(\e), &
\end{flalign}
\begin{flalign}
	{\cal T}^{(8,\left[1\times 1\right])}_{q\bar{q}g}\Big|_{N^0} = &
	+\frac{1}{\e^6}\left(-\frac{5}{8}\right)
	+\frac{1}{\e^5}\left(-\frac{179}{48}\right)
	+\frac{1}{\e^4}\left(-\frac{769}{48}+\frac{199}{288}\pi^2\right)
	\nonumber &\\&
	+\frac{1}{\e^3}\left(-\frac{6347}{96}+\frac{2375}{576}\pi^2+\frac{137}{8}\zeta_{3}\right)
	\nonumber &\\&
	+\frac{1}{\e^2}\left(-\frac{25441}{96}+\frac{9997}{576}\pi^2+\frac{12457}{144}\zeta_{3}+\frac{6817}{34560}\pi^4\right)
	\nonumber &\\&
	+\frac{1}{\e}\left(-\frac{207427}{192}+\frac{42641}{576}\pi^2+\frac{18713}{48}\zeta_{3}+\frac{37817}{69120}\pi^4-\frac{7411}{288}\pi^2\zeta_{3}+\frac{1903}{8}\zeta_{5}\right)
	\nonumber &\\&
	-\frac{1748021}{384}+\frac{358529}{1152}\pi^2+\frac{84229}{48}\zeta_{3}+\frac{248171}{69120}\pi^4-\frac{222493}{1728}\pi^2\zeta_{3}
	\nonumber &\\&
	+\frac{88921}{80}\zeta_{5}+\frac{607987}{2903040}\pi^6-\frac{5805}{16}\zeta_{3}^2 + \mathcal{O}(\e), &
\end{flalign}
\begin{flalign}
	{\cal T}^{(8,\left[1\times 1\right])}_{q\bar{q}g}\Big|_{1/N^2} = &
	+\frac{1}{\e^6}\left(\frac{1}{4}\right)
	+\frac{1}{\e^5}\left(\frac{9}{8}\right)
	+\frac{1}{\e^4}\left(\frac{83}{16}-\frac{13}{48}\pi^2\right)
	+\frac{1}{\e^3}\left(\frac{667}{32}-\frac{35}{32}\pi^2-\frac{85}{12}\zeta_{3}\right)
	\nonumber &\\&
	+\frac{1}{\e^2}\left(\frac{2607}{32}-\frac{959}{192}\pi^2-\frac{217}{8}\zeta_{3}-\frac{41}{384}\pi^4\right)
	\nonumber &\\&
	+\frac{1}{\e}\left(\frac{61319}{192}-\frac{7895}{384}\pi^2-\frac{6109}{48}\zeta_{3}-\frac{1349}{3840}\pi^4+\frac{1505}{144}\pi^2\zeta_{3}-\frac{2113}{20}\zeta_{5}\right)
	\nonumber &\\&
	+\frac{494285}{384}-\frac{32531}{384}\pi^2-\frac{53047}{96}\zeta_{3}-\frac{8137}{4608}\pi^4+\frac{3629}{96}\pi^2\zeta_{3}
	\nonumber &\\&
	-\frac{7191}{20}\zeta_{5}-\frac{161467}{1451520}\pi^6+\frac{1265}{8}\zeta_{3}^2 + \mathcal{O}(\e), &
\end{flalign}
\begin{flalign}
	{\cal T}^{(8,\left[1\times 1\right])}_{q\bar{q}g}\Big|_{\NF N} = &
	+\frac{1}{\e^5}\left(-\frac{5}{24}\right)
	+\frac{1}{\e^4}\left(-\frac{67}{72}\right)
	+\frac{1}{\e^3}\left(-\frac{47}{16}+\frac{13}{48}\pi^2\right)
	\nonumber &\\&
	+\frac{1}{\e^2}\left(-\frac{1481}{144}+\frac{107}{108}\pi^2+\frac{55}{18}\zeta_{3}\right)
	\nonumber &\\&
	+\frac{1}{\e}\left(-\frac{10385}{288}+\frac{32}{9}\pi^2+\frac{1265}{108}\zeta_{3}-\frac{41}{576}\pi^4\right)
	\nonumber &\\&
	-\frac{8305}{64}+\frac{22223}{1728}\pi^2+\frac{1517}{36}\zeta_{3}-\frac{10289}{51840}\pi^4-\frac{149}{36}\pi^2\zeta_{3}
	\nonumber &\\&
	+\frac{143}{6}\zeta_{5} + \mathcal{O}(\e), &
\end{flalign}
\begin{flalign}
	{\cal T}^{(8,\left[1\times 1\right])}_{q\bar{q}g}\Big|_{\NF/N} = &
	+\frac{1}{\e^5}\left(\frac{1}{6}\right)
	+\frac{1}{\e^4}\left(\frac{1}{2}\right)
	+\frac{1}{\e^3}\left(\frac{31}{16}-\frac{2}{9}\pi^2\right)
	+\frac{1}{\e^2}\left(\frac{79}{12}-\frac{5}{8}\pi^2-\frac{53}{18}\zeta_{3}\right)
	\nonumber &\\&
	+\frac{1}{\e}\left(\frac{1069}{48}-\frac{697}{288}\pi^2-\frac{91}{12}\zeta_{3}+\frac{19}{432}\pi^4\right)
	\nonumber &\\&
	+\frac{3653}{48}-\frac{25}{3}\pi^2-\frac{241}{8}\zeta_{3}+\frac{19}{144}\pi^4+\frac{493}{108}\pi^2\zeta_{3}-\frac{299}{10}\zeta_{5} + \mathcal{O}(\e), &
\end{flalign}
\begin{flalign}
	{\cal T}^{(8,\left[1\times 1\right])}_{q\bar{q}g}\Big|_{\NF^2} = &
	+\frac{1}{\e^4}\left(\frac{1}{36}\right)
	+\frac{1}{\e^3}\left(\frac{1}{24}\right)
	+\frac{1}{\e^2}\left(\frac{19}{144}-\frac{7}{432}\pi^2\right)
	\nonumber &\\&
	+\frac{1}{\e}\left(\frac{109}{288}-\frac{7}{288}\pi^2-\frac{25}{108}\zeta_{3}\right)
	\nonumber &\\&
	+\frac{71}{64}-\frac{133}{1728}\pi^2-\frac{25}{72}\zeta_{3}-\frac{71}{51840}\pi^4 + \mathcal{O}(\e). &
\end{flalign}
The three-gluon final state ${\cal T}^{(8)}_{ggg}$ is a genuine singlet contribution:
\begin{eqnarray}
	\label{eq:X32ansatz2}	
  {\cal T}^{(8)}_{ggg} &=&
  \left( N - \frac{1}{N}\right) {\cal T}^{(2)}_{q\bar q}
  \NFZ \left(N - \frac{4}{N}\right) \,{\cal T}^{(8)}_{ggg}\Big|_{\NFZ}\,,
\end{eqnarray}
with
\begin{flalign}
	{\cal T}^{(8)}_{ggg}\Big|_{\NFZ} = &
	-\frac{31}{12}+\frac{41}{144}\pi^2-\frac{8}{3}\zeta_{3}+\frac{7}{720}\pi^4-\frac{\pi^2\zeta_{3}}{6}+\frac{25}{6}\zeta_{5} + \mathcal{O}(\e)\,. &
\end{flalign}
In order to compute the renormalised ${\cal T}^{(8)}_{q\bar{q}g}$, we needed to
evaluate the master integrals $V_{5,a}$, $V_{5,b}$ and $V_8$ up to
weight 5, in the notation of~\cite{Gehrmann-DeRidder:2004ttg}. $V_{5,a}$ and
$V_{5,b}$ are provided in a closed form, whereas the expansion of $V_8$ is
truncated at weight 4~\cite{Gehrmann-DeRidder:2003pne}. We numerically computed
the master integral $V_8$ up to $\mathcal{O}(\e^2)$ with
\code{AMFlow}~\cite{Liu:2022chg}, and we reconstructed the analytic expression
with the PSLQ algorithm~\cite{pslq}. The result reads:
\begin{eqnarray}
	V_8 = S_{\Gamma,2} (q^2)^{-2-2\e} \Bigg[
	&-&\frac{5}{2\e^4}+\frac{9\pi^2}{2\e^2}+\frac{89\zeta_3}{\e}+\frac{13\pi^4}{180} +\e\left(-\frac{407\pi^2\zeta_3}{3} + 1135\zeta_5\right) \nonumber \\
	\hspace{2.5cm}&+&  \e^2\left(\frac{1451\pi^6}{2268} - 1181\zeta_3^2\right)+ {\cal O}(\e^3)\Bigg]\,,
\end{eqnarray}
with
\begin{equation}
	S_{\Gamma,2} = P_2\,\left( \frac{(4\pi)^\e}{16\pi^2\Gamma(1-\e)} \right)^2\,.
\end{equation}
As a validation of this result, we notice that the master integral denoted by
VVRR$_{22}$ in~\cite{Magerya:2019cvz} is related to $V_8$ by a trivial
multiplication with a one-loop bubble, as can be inferred by the definition of
the two integrals. We find complete agreement between the two expressions.

\subsection{Four-parton final states}
The processes contributing to the four-parton final state are
$\gamma^{*}\to q\bar{q}gg$, $\gamma^{*}\to q\bar{q}q'\bar{q}'$ and
$\gamma^{*}\to q\bar{q}q\bar{q}$, where $q'$ denotes a quark of a flavour
different to that of $q$.
The amplitude up to one loop is
\begin{equation}
|{\cal M}\rangle_{q\bar q ij} = \sqrt{4\pi\alpha}e_q  \,
4\pi\alpha_s \,
\left[ |{\cal M}^{(0)}\rangle_{q\bar q ij}
+ \left(\frac{\alpha_s}{2\pi}\right) |{\cal M}^{(1)}\rangle_{q\bar q ij}
+ {\cal O}(\alpha_s^2) \right]\,,
\end{equation}
with $ij = q'\bar q', q\bar q, gg$.
The perturbative expansions of the squared amplitudes summed over spins, colours and quark flavours are
\begin{eqnarray}
\langle{\cal M}|{\cal M}\rangle_{q\bar q ij} &=& \sum |{\cal M}(\gamma^* \to q\bar q ij)|^2 \nonumber\\
&=& 4\pi\alpha\sum_q e_q^2\,64\pi^4
\Bigg[\left(\frac{\alpha_s}{2\pi}\right)^2
\langle{\cal M}^{(0)}|{\cal M}^{(0)}\rangle_{q\bar q ij} 
\nonumber \\
&& \hspace{1.3cm}
+ \left(\frac{\alpha_s}{2\pi}\right)^3
\left(2\,\Re\big[\langle{\cal M}^{(1)}|{\cal M}^{(0)}\rangle_{q\bar q ij} \big]
\right) + {\cal O}(\alpha_s^4) \Bigg]. 
\end{eqnarray}
We define
\begin{eqnarray}
	{\cal T}^{(6)}_{q\bar q ij} 
	 &=& 64\pi^4 \int \d \Phi_4 \,
	\langle{\cal M}^{(0)}|{\cal M}^{(0)}\rangle_{q\bar q ij}\,, \\
	{\cal T}^{(8)}_{q\bar q ij} 
	 &=& 64\pi^4 \int \d \Phi_4 \,
	2\,\Re\big[\langle{\cal M}^{(1)}|{\cal M}^{(0)}\rangle_{q\bar q ij}\big]. 
\end{eqnarray}
The expressions for ${\cal T}^{(6)}_{q\bar q ij}$ are given in equations~(4.49),~(4.51) and ~(4.53) of \cite{Gehrmann-DeRidder:2004ttg} and in Appendix~\ref{app:A}. The colour decomposition of
${\cal T}^{(8)}_{q \bar q gg}$ reads
\begin{eqnarray}
	\label{eq:X41ansatz1}	
	{\cal T}^{(8)}_{q \bar q gg} &=& \left( N - \frac{1}{N}\right) {\cal T}^{(2)}_{q\bar q}\, \Bigg[
	N^2 {\cal T}^{(8)}_{q \bar q gg}\Big|_{N^2} + {\cal T}^{(8)}_{q \bar q gg}\Big|_{N^0}
	+ \frac{1}{N^2} {\cal T}^{(8)}_{q \bar q gg}\Big|_{1/N^2}
	\nonumber \\ &&                       
	+ \NF N\, {\cal T}^{(8)}_{q \bar q gg}\Big|_{\NF N}
	+ \frac{\NF}{N}\,{\cal T}^{(8)}_{q \bar q gg}\Big|_{\NF/N}
	+ \NFZ \left(N - \frac{4}{N}\right) \,{\cal T}^{(8)}_{q \bar q gg}\Big|_{\NFZ}
	\Bigg]\,.
\end{eqnarray}
The results for the coefficients are
\begin{flalign}
	{\cal T}^{(8)}_{q\bar{q}gg}\Big|_{N^2} = &
	+\frac{1}{\e^6}\left(-\frac{41}{36}\right)
	+\frac{1}{\e^5}\left(-\frac{311}{36}\right)
	+\frac{1}{\e^4}\left(-\frac{54325}{1296}+\frac{1151}{432}\pi^2\right)
	\nonumber &\\&        
	+\frac{1}{\e^3}\left(-\frac{1590017}{7776}+\frac{23519}{1296}\pi^2+\frac{380}{9}\zeta_{3}\right)
	\nonumber &\\&
	+\frac{1}{\e^2}\left(-\frac{7646353}{7776}+\frac{1461895}{15552}\pi^2+\frac{66593}{216}\zeta_{3}-\frac{16537}{10368}\pi^4\right)
	\nonumber &\\&
	+\frac{1}{\e}\bigg(-\frac{84015367}{17496}+\frac{44125165}{93312}\pi^2+\frac{2192809}{1296}\zeta_{3}
	\nonumber &\\& \phantom{+\frac{1}{\e}\bigg(\;\;\;}
        -\frac{7973}{960}\pi^4-\frac{7435}{72}\pi^2\zeta_{3}+\frac{40319}{90}\zeta_{5}\bigg)
	\nonumber &\\&
	-\frac{20052623335}{839808}+\frac{163823405}{69984}\pi^2+\frac{22820177}{2592}\zeta_{3}-\frac{4992721}{124416}\pi^4
	\nonumber &\\&        
        -\frac{22493}{32}\pi^2\zeta_{3}+\frac{1335263}{360}\zeta_{5}+\frac{4433837}{13063680}\pi^6-\frac{64345}{72}\zeta_{3}^2 + \mathcal{O}(\e)\,, &
\end{flalign}
\begin{flalign}
	{\cal T}^{(8)}_{q\bar{q}gg}\Big|_{N^0} = &
	+\frac{1}{\e^6}\left(\frac{3}{2}\right)
	+\frac{1}{\e^5}\left(\frac{217}{24}\right)
	+\frac{1}{\e^4}\left(\frac{389}{9}-\frac{65}{18}\pi^2\right)
	\nonumber &\\&        
	+\frac{1}{\e^3}\left(\frac{177335}{864}-\frac{17189}{864}\pi^2-\frac{761}{12}\zeta_{3}\right)
	\nonumber &\\&
	+\frac{1}{\e^2}\left(\frac{4999865}{5184}-\frac{128357}{1296}\pi^2-\frac{12755}{36}\zeta_{3}+\frac{16579}{8640}\pi^4\right)
	\nonumber &\\&
	+\frac{1}{\e}\bigg(\frac{143110091}{31104}-\frac{14956337}{31104}\pi^2-\frac{65791}{36}\zeta_{3}
	\nonumber &\\& \phantom{+\frac{1}{\e}\bigg(\;\;\;}
        +\frac{310769}{34560}\pi^4+\frac{23987}{144}\pi^2\zeta_{3}-\frac{51389}{60}\zeta_{5}\bigg)
	\nonumber &\\&
	+\frac{4179822425}{186624}-\frac{431838125}{186624}\pi^2-\frac{7979983}{864}\zeta_{3}+\frac{62213}{1440}\pi^4
	\nonumber &\\&        
        +\frac{190001}{216}\pi^2\zeta_{3}-\frac{856933}{180}\zeta_{5}-\frac{719437}{1088640}\pi^6+\frac{19345}{12}\zeta_{3}^2 + \mathcal{O}(\e)\,, &
\end{flalign}
\begin{flalign}
	{\cal T}^{(8)}_{q\bar{q}gg}\Big|_{1/N^2} = &
	+\frac{1}{\e^6}\left(-\frac{1}{2}\right)
	+\frac{1}{\e^5}\left(-\frac{9}{4}\right)
	+\frac{1}{\e^4}\left(-\frac{91}{8}+\frac{29}{24}\pi^2\right)
	\nonumber &\\&        
	+\frac{1}{\e^3}\left(-\frac{1683}{32}+\frac{83}{16}\pi^2+\frac{137}{6}\zeta_{3}\right)
	\nonumber &\\&
	+\frac{1}{\e^2}\left(-\frac{1937}{8}+\frac{843}{32}\pi^2+\frac{373}{4}\zeta_{3}-\frac{169}{320}\pi^4\right)
	\nonumber &\\&
	+\frac{1}{\e}\left(-\frac{54331}{48}+\frac{142745}{1152}\pi^2+\frac{965}{2}\zeta_{3}-\frac{13073}{5760}\pi^4-\frac{1463}{24}\pi^2\zeta_{3}+\frac{10469}{30}\zeta_{5}\right)
	\nonumber &\\&
	-\frac{2082919}{384}+\frac{168199}{288}\pi^2+\frac{222847}{96}\zeta_{3}-\frac{14011}{1280}\pi^4
	\nonumber &\\&        
        -\frac{3885}{16}\pi^2\zeta_{3}+\frac{56609}{40}\zeta_{5}+\frac{51599}{311040}\pi^6-\frac{7667}{12}\zeta_{3}^2 + \mathcal{O}(\e)\,, &
\end{flalign}
\begin{flalign}
	{\cal T}^{(8)}_{q\bar{q}gg}\Big|_{\NF N} = &
	+\frac{1}{\e^5}\left(\frac{1}{2}\right)
	+\frac{1}{\e^4}\left(\frac{65}{36}\right)
	+\frac{1}{\e^3}\left(\frac{869}{108}-\frac{13}{18}\pi^2\right)
	+\frac{1}{\e^2}\left(\frac{14459}{432}-\frac{589}{216}\pi^2-\frac{71}{6}\zeta_{3}\right)
	\nonumber &\\&
	+\frac{1}{\e}\left(\frac{1084631}{7776}-\frac{15943}{1296}\pi^2-\frac{1327}{27}\zeta_{3}+\frac{373}{2160}\pi^4\right)
	\nonumber &\\&
	+\frac{27302353}{46656}-\frac{200555}{3888}\pi^2-\frac{18365}{81}\zeta_{3}+\frac{5207}{12960}\pi^4
	\nonumber &\\&
        +\frac{1891}{108}\pi^2\zeta_{3}-\frac{887}{10}\zeta_{5} + \mathcal{O}(\e)\,, &
\end{flalign}
\begin{flalign}
	{\cal T}^{(8)}_{q\bar{q}gg}\Big|_{\NF/N} = &
 	+\frac{1}{\e^5}\left(-\frac{1}{3}\right)
	+\frac{1}{\e^4}\left(-1\right)
	+\frac{1}{\e^3}\left(-\frac{13}{3}+\frac{\pi^2}{2}\right)
	+\frac{1}{\e^2}\left(-\frac{845}{48}+\frac{3}{2}\pi^2+\frac{80}{9}\zeta_{3}\right)
	\nonumber &\\&
	+\frac{1}{\e}\left(-\frac{2307}{32}+\frac{473}{72}\pi^2+\frac{80}{3}\zeta_{3}-\frac{17}{216}\pi^4\right)
	\nonumber &\\&
	-\frac{57305}{192}+\frac{7735}{288}\pi^2+\frac{4295}{36}\zeta_{3}-\frac{17}{72}\pi^4-\frac{121}{9}\pi^2\zeta_{3}+\frac{424}{5}\zeta_{5} + \mathcal{O}(\e)\,, &
\end{flalign}

\begin{flalign}
	{\cal T}^{(8)}_{q\bar{q}gg}\Big|_{\NFZ} = &
	+\frac{139}{12}-\frac{49}{72}\pi^2-\frac{5}{4}\zeta_{3}-\frac{7}{720}\pi^4+\frac{2}{3}\pi^2\zeta_{3}-10  \zeta_{5} + \mathcal{O}(\e)\,. &
\end{flalign}
The colour decomposition of ${\cal T}^{(8)}_{q \bar q q'\bar{q}'}$ is given by
\begin{eqnarray}
	\label{eq:X41ansatz2}	
	{\cal T}^{(8)}_{q \bar q q'\bar{q}'} &=& \left( N - \frac{1}{N}\right) {\cal T}^{(2)}_{q\bar q}\, \Bigg[
	(\NF-1) N\, {\cal T}^{(8)}_{q \bar q q'\bar{q}'}\Big|_{\NF N}
	+ \frac{(\NF-1)}{N}\,{\cal T}^{(8)}_{q \bar q q'\bar{q}'}\Big|_{\NF/N} \nonumber \\ &&
	+ (\NF-1)\NF \,{\cal T}^{(8)}_{q \bar q q'\bar{q}'}\Big|_{\NF^2}
	+ (\NFZ-1) \left(N - \frac{4}{N}\right) \,{\cal T}^{(8)}_{q \bar q q'\bar{q}'}\Big|_{\NFZ}
	\Bigg]\,,
\end{eqnarray}
where we have used $\NF-1$ and $\NFZ-1$ because of the different flavour of the two final-state quark lines. The coefficients above read:
\begin{flalign}
	{\cal T}^{(8)}_{q\bar{q}q'\bar{q}'}\Big|_{\NF N} = &
	+\frac{1}{\e^5}\left(\frac{13}{108}\right)
	+\frac{1}{\e^4}\left(\frac{679}{648}\right)
	+\frac{1}{\e^3}\left(\frac{11249}{1944}-\frac{425}{1296}\pi^2\right)
	\nonumber &\\&
	+\frac{1}{\e^2}\left(\frac{185695}{5832}-\frac{20297}{7776}\pi^2-\frac{265}{36}\zeta_{3}\right)
	\nonumber &\\&
	+\frac{1}{\e}\left(\frac{1554361}{8748}-\frac{344569}{23328}\pi^2-\frac{1457}{24}\zeta_{3}+\frac{3239}{51840}\pi^4\right)
	\nonumber &\\&
	+\frac{210844381}{209952}-\frac{1432307}{17496}\pi^2-\frac{8225}{24}\zeta_{3}+\frac{64091}{311040}\pi^4
	\nonumber &\\&	
	+\frac{8261}{432}\pi^2\zeta_{3}-\frac{9121}{60}\zeta_{5} + \mathcal{O}(\e)\,, &
\end{flalign}
\begin{flalign}
	{\cal T}^{(8)}_{q\bar{q}q'\bar{q}'}\Big|_{\NF/N} = &
	+\frac{1}{\e^5}\left(-\frac{11}{108}\right)
	+\frac{1}{\e^4}\left(-\frac{425}{648}\right)
	+\frac{1}{\e^3}\left(-\frac{7627}{1944}+\frac{133}{432}\pi^2\right)
	\nonumber &\\&
	+\frac{1}{\e^2}\left(-\frac{131273}{5832}+\frac{5047}{2592}\pi^2+\frac{889}{108}\zeta_{3}\right)
	\nonumber &\\&
	+\frac{1}{\e}\left(-\frac{2276611}{17496}+\frac{89357}{7776}\pi^2+\frac{34279}{648}\zeta_{3}-\frac{125}{10368}\pi^4\right)
	\nonumber &\\&
	-\frac{159423833}{209952}+\frac{1510867}{23328}\pi^2+\frac{606965}{1944}\zeta_{3}+\frac{17861}{311040}\pi^4
	\nonumber &\\&        
        -\frac{10375}{432}\pi^2\zeta_{3}+\frac{34661}{180}\zeta_{5} + \mathcal{O}(\e)\,, &
\end{flalign}
\begin{flalign}
	{\cal T}^{(8)}_{q\bar{q}q'\bar{q}'}\Big|_{\NF^2} = &
	+\frac{1}{\e^4}\left(-\frac{7}{162}\right)
	+\frac{1}{\e^3}\left(-\frac{79}{486}\right)
	+\frac{1}{\e^2}\left(-\frac{97}{162}+\frac{\pi^2}{18}\right)
        \nonumber &\\&
	+\frac{1}{\e}\left(-\frac{3613}{2187}+\frac{73}{648}\pi^2+\frac{76}{81}\zeta_{3}\right)
	\nonumber &\\&
	+\frac{4661}{26244}-\frac{523}{3888}\pi^2-\frac{247}{486}\zeta_{3}+\frac{41}{38880}\pi^4 + \mathcal{O}(\e)\,, &
\end{flalign}
\begin{flalign}
  {\cal T}^{(8)}_{q\bar{q}q'\bar{q}'}\Big|_{\NFZ} = &
  +\frac{1}{\e}\left(-\frac{7 }{3}+\frac{\pi^2 }{9}+\frac{\zeta_{3} }{2}+\frac{\pi^4 }{135}\right)
\nonumber &\\&
-\frac{2335 }{72}+\frac{119 \pi^2 }{108}+\frac{19 \zeta_{3} }{3}+\frac{961 \pi^4 }{12960}-\frac{14 \pi^2\zeta_{3} }{9}+26  \zeta_{5} + \mathcal{O}(\e)\,. &
\end{flalign}
Finally, the colour decomposition of ${\cal T}^{(8)}_{q \bar q q\bar{q}}$ is
\begin{eqnarray}
	\label{eq:X41ansatz3}	
	\hspace{-0.3cm}{\cal T}^{(8)}_{q \bar q q\bar{q}} &=& \left( N - \frac{1}{N}\right) {\cal T}^{(2)}_{q\bar q}\, \Bigg[                    
	N\, {\cal T}^{(8)}_{q \bar q q'\bar{q}'}\Big|_{\NF N}
	+ \frac{1}{N}\,{\cal T}^{(8)}_{q \bar q q'\bar{q}'}\Big|_{\NF/N}
        + \NF \,{\cal T}^{(8)}_{q \bar q q'\bar{q}'}\Big|_{\NF^2}
	\nonumber \\ &&                                                      
	+ \left(N - \frac{4}{N}\right) \,{\cal T}^{(8)}_{q \bar q q'\bar{q}'}\Big|_{\NFZ}
	+ {\cal T}^{(8)}_{q \bar q q\bar{q}}\Big|_{N^0}
	+ \frac{1}{N^2} {\cal T}^{(8)}_{q \bar q q\bar{q}}\Big|_{1/N^2} 
	+ \frac{\NF}{N} {\cal T}^{(8)}_{q \bar q q\bar{q}}\Big|_{\NF/N}
	\Bigg],
\end{eqnarray}
where the first three terms are identical to those
in~\eqref{eq:X41ansatz2}. The
factors $\NF$ and $\NFZ$ are correctly restored in the sum of the
$q\bar{q}q'\bar{q}'$ and $q\bar{q}q\bar{q}$ sub-channels. The new terms appearing in the same-flavour case are
\begin{flalign}
	{\cal T}^{(8)}_{q\bar{q}q\bar{q}}\Big|_{N^0} = &
	+\frac{1}{\e^3}\left(-\frac{65}{48}+\frac{5}{24}\pi^2-\frac{5}{6}\zeta_{3}\right)
	+\frac{1}{\e^2}\left(-\frac{3527}{144}+\frac{275}{144}\pi^2+\frac{80}{9}\zeta_{3}-\frac{101}{1080}\pi^4\right)
	\nonumber &\\&
	+\frac{1}{\e}\left(-\frac{217805}{864}+\frac{23891}{1728}\pi^2+\frac{10573}{108}\zeta_{3}+\frac{281}{6480}\pi^4+\frac{169}{72}\pi^2\zeta_{3}-\frac{117}{2}\zeta_{5}\right)
	\nonumber &\\&
	-\frac{325597}{162}+\frac{280337}{2592}\pi^2+\frac{97657}{162}\zeta_{3}+\frac{171733}{77760}\pi^4
	\nonumber &\\&        
        -\frac{6265}{216}\pi^2\zeta_{3}+\frac{4403}{18}\zeta_{5}-\frac{8171}{90720}\pi^6+\frac{110}{3}\zeta_{3}^2 + \mathcal{O}(\e)\,, &
\end{flalign}
\begin{flalign}
	{\cal T}^{(8)}_{q\bar{q}q\bar{q}}\Big|_{1/N^2} = &
	+\frac{1}{\e^3}\left(\frac{65}{48}-\frac{5}{24}\pi^2+\frac{5}{6}\zeta_{3}\right)
	+\frac{1}{\e^2}\left(\frac{185}{8}-\frac{79}{48}\pi^2-9 \zeta_{3}+\frac{7}{90}\pi^4\right)
	\nonumber &\\&
	+\frac{1}{\e}\left(\frac{45953}{192}-\frac{7435}{576}\pi^2-\frac{935}{12}\zeta_{3}-\frac{43}{720}\pi^4-\frac{229}{72}\pi^2\zeta_{3}+\frac{277}{6}\zeta_{5}\right)
	\nonumber &\\&
	+\frac{251839}{128}-\frac{30857}{288}\pi^2-\frac{24691}{48}\zeta_{3}-\frac{5089}{4320}\pi^4
	\nonumber &\\&
        +\frac{281}{12}\pi^2\zeta_{3}-\frac{421}{2}\zeta_{5}-\frac{11}{336}\pi^6-76 \zeta_{3}^2 + \mathcal{O}(\e)\,, &
\end{flalign}
\begin{flalign}
	{\cal T}^{(8)}_{q\bar{q}q\bar{q}}\Big|_{\NF/N} = &
	+\frac{1}{\e^2}\left(\frac{13}{72}-\frac{\pi^2}{36}+\frac{\zeta_{3}}{9}\right)
	+\frac{1}{\e}\left(\frac{19}{108}+\frac{11}{216}\pi^2-\frac{28}{27}\zeta_{3}+\frac{17}{3240}\pi^4\right)
	\nonumber &\\&
	-\frac{3437}{162}+\frac{247}{162}\pi^2+\frac{2125}{324}\zeta_{3}-\frac{1249}{19440}\pi^4+\frac{8}{27}\pi^2\zeta_{3}-\frac{19}{9}\zeta_{5} + \mathcal{O}(\e)\,. &
\end{flalign}

\subsection{Five-parton final states}
The processes contributing to the five-parton final state are
$\gamma^{*}\to q\bar{q}ggg$, $\gamma^{*}\to q\bar{q}q'\bar{q}'g$ and
$\gamma^{*}\to q\bar{q}q\bar{q}g$, and the tree-level amplitudes read
\begin{equation}
	|{\cal M}\rangle_{q\bar q ijk} = \sqrt{4\pi\alpha}e_q  \,
	(4\pi\alpha_s)^{3/2} \,
	\left[ |{\cal M}^{(0)}\rangle_{q\bar q ijk} 
	+ {\cal O}(\alpha_s) \right]\,,
\end{equation}
with $ijk = gg g,q'\bar q' g, q\bar q g$.  The squared amplitude, summed over
spins, colours and quark flavours is
\begin{eqnarray}
	\langle{\cal M}|{\cal M}\rangle_{q\bar q ijk} &=& \sum |{\cal M}(\gamma^* \to q\bar q ijk)|^2 \nonumber \\
	&=& 4\pi\alpha\sum_q e_q^2\, 512\pi^6\,
	\Bigg[\left(\frac{\alpha_s}{2\pi}\right)^3\,
	\langle{\cal M}^{(0)}|{\cal M}^{(0)}\rangle_{q\bar q ijk} 
	+ {\cal O}(\alpha_s^4) \Bigg]\,.
\end{eqnarray}
We define
\begin{eqnarray}
	{\cal T}^{(8)}_{q\bar q ijk} 
	 &=& 512\pi^8 \int \d \Phi_5 \,
	\langle{\cal M}^{(0)}|{\cal M}^{(0)}\rangle_{q\bar q ijk}\,.
\end{eqnarray}
The colour decomposition for ${\cal T}^{(8)}_{q \bar q ggg}$ reads:
\begin{eqnarray}
	\label{eq:X50ansatz1}
	{\cal T}^{(8)}_{q \bar q ggg} &=& \left( N - \frac{1}{N}\right) {\cal T}^{(2)}_{q\bar q}\, \Bigg[
	N^2 {\cal T}^{(8)}_{q \bar q ggg}\Big|_{N^2} + {\cal T}^{(8)}_{q \bar q ggg}\Big|_{N^0}
	+ \frac{1}{N^2} {\cal T}^{(8)}_{q \bar q ggg}\Big|_{1/N^2}
	\Bigg]\,,
\end{eqnarray}
with
\begin{flalign}
	{\cal T}^{(8)}_{q\bar{q}ggg}\Big|_{N^2} = &
	+\frac{1}{\e^6}\left(\frac{1}{2}\right)
	+\frac{1}{\e^5}\left(\frac{331}{108}\right)
	+\frac{1}{\e^4}\left(\frac{11843}{648}-\frac{31}{24}\pi^2\right)
	\nonumber &\\&        
	+\frac{1}{\e^3}\left(\frac{259867}{2592}-\frac{10745}{1296}\pi^2-\frac{439}{18}\zeta_{3}\right)
	\nonumber &\\&
	+\frac{1}{\e^2}\left(\frac{6302057}{11664}-\frac{394223}{7776}\pi^2-\frac{6239}{36}\zeta_{3}+\frac{21853}{25920}\pi^4\right)
	\nonumber &\\&
	+\frac{1}{\e}\Bigg(\frac{815913157}{279936}-\frac{26347837}{93312}\pi^2-\frac{181151}{162}\zeta_{3}
	\nonumber &\\& \phantom{+\frac{1}{\e}\Bigg(\;\;}
        +\frac{75767}{17280}\pi^4+\frac{13993}{216}\pi^2\zeta_{3}-\frac{10946}{45}\zeta_{5}\bigg)
	\nonumber &\\&
	+\frac{736904809}{46656}-\frac{107045579}{69984}\pi^2-\frac{49920557}{7776}\zeta_{3}+\frac{7130357}{311040}\pi^4
	\nonumber &\\&
        +\frac{67895}{144}\pi^2\zeta_{3}-\frac{103894}{45}\zeta_{5}-\frac{93257}{1306368}\pi^6+\frac{7861}{12}\zeta_{3}^2 + \mathcal{O}(\e)\,, &
\end{flalign}
\begin{flalign}
	{\cal T}^{(8)}_{q\bar{q}ggg}\Big|_{N^0} = &
	+\frac{1}{\e^6}\left(-\frac{7}{12}\right)
	+\frac{1}{\e^5}\left(-\frac{37}{12}\right)
	+\frac{1}{\e^4}\left(-\frac{1255}{72}+\frac{25}{16}\pi^2\right)
	\nonumber &\\&        
	+\frac{1}{\e^3}\left(-\frac{39895}{432}+\frac{76}{9}\pi^2+\frac{63}{2}\zeta_{3}\right)
	\nonumber &\\&
	+\frac{1}{\e^2}\left(-\frac{78673}{162}+\frac{20903}{432}\pi^2+\frac{12907}{72}\zeta_{3}-\frac{15811}{17280}\pi^4\right)
	\nonumber &\\&
	+\frac{1}{\e}\left(-\frac{40021591}{15552}+\frac{335677}{1296}\pi^2+\frac{458257}{432}\zeta_{3}-\frac{15115}{3456}\pi^4-\frac{1541}{18}\pi^2\zeta_{3}+\frac{5774}{15}\zeta_{5}\right)
	\nonumber &\\&
	-\frac{320911123}{23328}+\frac{42522607}{31104}\pi^2+\frac{15074377}{2592}\zeta_{3}-\frac{2348393}{103680}\pi^4
	\nonumber &\\&        
        -\frac{142223}{288}\pi^2\zeta_{3}+\frac{890561}{360}\zeta_{5}+\frac{916243}{4354560}\pi^6-\frac{21473}{24}\zeta_{3}^2 + \mathcal{O}(\e)\,, &
\end{flalign}
\begin{flalign}
	{\cal T}^{(8)}_{q\bar{q}ggg}\Big|_{1/N^2} = &
	+\frac{1}{\e^6}\left(\frac{1}{6}\right)
	+\frac{1}{\e^5}\left(\frac{3}{4}\right)
	+\frac{1}{\e^4}\left(\frac{33}{8}-\frac{11}{24}\pi^2\right)
	+\frac{1}{\e^3}\left(\frac{687}{32}-\frac{33}{16}\pi^2-\frac{59}{6}\zeta_{3}\right)
	\nonumber &\\&
	+\frac{1}{\e^2}\left(\frac{1787}{16}-\frac{1099}{96}\pi^2-\frac{177}{4}\zeta_{3}+\frac{659}{2880}\pi^4\right)
	\nonumber &\\&
	+\frac{1}{\e}\left(\frac{225217}{384}-\frac{69133}{1152}\pi^2-\frac{1997}{8}\zeta_{3}+\frac{1973}{1920}\pi^4+\frac{1981}{72}\pi^2\zeta_{3}-\frac{1451}{10}\zeta_{5}\right)
	\nonumber &\\&
	+\frac{597437}{192}-\frac{180415}{576}\pi^2-\frac{126785}{96}\zeta_{3}+\frac{187531}{34560}\pi^4
	\nonumber &\\&        
        +\frac{5917}{48}\pi^2\zeta_{3}-\frac{26103}{40}\zeta_{5}-\frac{197047}{2177280}\pi^6+\frac{1831}{6}\zeta_{3}^2 + \mathcal{O}(\e)\,. &
\end{flalign}
The colour decomposition for ${\cal T}^{(8)}_{q \bar q q'\bar{q}' g}$ reads:
\begin{eqnarray}
	\label{eq:X50ansatz2}
	{\cal T}^{(8)}_{q \bar q q'\bar{q}'g} &=& \left( N - \frac{1}{N}\right) {\cal T}^{(2)}_{q\bar q}\, \Bigg[                       
	+ (\NF-1) N\, {\cal T}^{(8)}_{q \bar q q'\bar{q}'g}\Big|_{\NF N}
	+ \frac{(\NF-1)}{N}\,{\cal T}^{(8)}_{q \bar q q'\bar{q}'g}\Big|_{\NF/N} \nonumber \\ &&
	+ (\NFZ-1) \left(N - \frac{4}{N}\right) \,{\cal T}^{(8)}_{q \bar q q'\bar{q}'g}\Big|_{\NFZ}
	\Bigg]\,,
\end{eqnarray}
with
\begin{flalign}
	{\cal T}^{(8)}_{q\bar{q}q'\bar{q}'g}\Big|_{\NF N} = &
	+\frac{1}{\e^5}\left(-\frac{7}{54}\right)
	+\frac{1}{\e^4}\left(-\frac{101}{108}\right)
	+\frac{1}{\e^3}\left(-\frac{11651}{1944}+\frac{247}{648}\pi^2\right)
	\nonumber &\\&
	+\frac{1}{\e^2}\left(-\frac{426175}{11664}+\frac{11227}{3888}\pi^2+\frac{493}{54}\zeta_{3}\right)
	\nonumber &\\&
	+\frac{1}{\e}\left(-\frac{5169679}{23328}+\frac{437695}{23328}\pi^2+\frac{24599}{324}\zeta_{3}-\frac{3613}{25920}\pi^4\right)
	\nonumber &\\&
	-\frac{563396717}{419904}+\frac{15881671}{139968}\pi^2+\frac{328225}{648}\zeta_{3}-\frac{17281}{31104}\pi^4
	\nonumber &\\&        
        -\frac{205}{8}\pi^2\zeta_{3}+\frac{13757}{90}\zeta_{5} + \mathcal{O}(\e)\,, &
\end{flalign}
\begin{flalign}
	{\cal T}^{(8)}_{q\bar{q}q'\bar{q}'g}\Big|_{\NF/N} = &
	+\frac{1}{\e^5}\left(\frac{11}{108}\right)
	+\frac{1}{\e^4}\left(\frac{425}{648}\right)
	+\frac{1}{\e^3}\left(\frac{15821}{3888}-\frac{47}{144}\pi^2\right)
	\nonumber &\\&
	+\frac{1}{\e^2}\left(\frac{573287}{23328}-\frac{1829}{864}\pi^2-\frac{979}{108}\zeta_{3}\right)
	\nonumber &\\&
	+\frac{1}{\e}\left(\frac{20849357}{139968}-\frac{67769}{5184}\pi^2-\frac{38797}{648}\zeta_{3}+\frac{1777}{51840}\pi^4\right)
	\nonumber &\\&
	+\frac{757887911}{839808}-\frac{2403131}{31104}\pi^2-\frac{1436257}{3888}\zeta_{3}+\frac{36571}{311040}\pi^4
	\nonumber &\\&        
        +\frac{1261}{48}\pi^2\zeta_{3}-\frac{34651}{180}\zeta_{5} + \mathcal{O}(\e)\,, &
\end{flalign}
\begin{flalign}
	{\cal T}^{(8)}_{q\bar{q}q'\bar{q}'g}\Big|_{\NFZ} = &
	+\frac{1}{\e}\left(\frac{7}{3}-\frac{\pi^2}{9}-\frac{\zeta_{3}}{2}-\frac{\pi^4}{135}\right)
	\nonumber &\\&
	+\frac{977}{36}-\frac{299}{432}\pi^2-\frac{25}{6}\zeta_{3}-\frac{979}{12960}\pi^4+\frac{25}{18}\pi^2\zeta_{3}-\frac{151  \zeta_{5}}{6} + \mathcal{O}(\e)\,. &
\end{flalign}
Finally, the colour decomposition for ${\cal T}^{(8)}_{q \bar q q\bar{q} g}$
reads:
\begin{eqnarray}
\label{eq:X50ansatz3}
	{\cal T}^{(8)}_{q \bar q q \bar q g} &=& \left( N - \frac{1}{N}\right) {\cal T}^{(2)}_{q\bar q}\, \Bigg[
	N\, {\cal T}^{(8)}_{q \bar q q' \bar{q}' g}\Big|_{\NF N}
	+ \frac{1}{N}\, {\cal T}^{(8)}_{q \bar q q' \bar{q}' g}\Big|_{\NF/N}
	\nonumber \\ &&
        + \left(N - \frac{4}{N}\right) \,{\cal T}^{(8)}_{q \bar q q'\bar{q}'g}\Big|_{\NFZ} 
	+ \,{\cal T}^{(8)}_{q \bar q q \bar q g}\Big|_{N^0}
	+ \frac{1}{N^2}\,{\cal T}^{(8)}_{q \bar q q \bar q g}\Big|_{1/N^2}
	\Bigg]\,,
\end{eqnarray}
and analogously to the four-parton final state case, the first three
contributions in~\eqref{eq:X50ansatz3} are inherited
from~\eqref{eq:X50ansatz2}. The new terms are given by:
\begin{flalign}
	{\cal T}^{(8)}_{q\bar{q}q\bar{q}g}\Big|_{N^0} = &
	+\frac{1}{\e^3}\left(\frac{65}{48}-\frac{5}{24}\pi^2+\frac{5}{6}\zeta_{3}\right)
	+\frac{1}{\e^2}\left(\frac{47}{2}-\frac{253}{144}\pi^2-\frac{19}{2}\zeta_{3}+\frac{101}{1080}\pi^4\right)
	\nonumber &\\&
	+\frac{1}{\e}\left(\frac{48383}{192}-\frac{2761}{192}\pi^2-\frac{731}{8}\zeta_{3}-\frac{13}{240}\pi^4-\frac{19}{8}\pi^2\zeta_{3}+\frac{176}{3}\zeta_{5}\right)
	\nonumber &\\&
	+\frac{411893}{192}-\frac{69037}{576}\pi^2-\frac{10409}{16}\zeta_{3}-\frac{14657}{8640}\pi^4
	\nonumber &\\&        
        +\frac{731}{24}\pi^2\zeta_{3}-\frac{1941}{8}\zeta_{5}+\frac{167}{1701}\pi^6-\frac{511}{12}\zeta_{3}^2 + \mathcal{O}(\e)\,, &
\end{flalign}
\begin{flalign}
	{\cal T}^{(8)}_{q\bar{q}q\bar{q}g}\Big|_{1/N^2} = &
	+\frac{1}{\e^3}\left(-\frac{65}{48}+\frac{5}{24}\pi^2-\frac{5}{6}\zeta_{3}\right)
	+\frac{1}{\e^2}\left(-\frac{185}{8}+\frac{79}{48}\pi^2+9 \zeta_{3}-\frac{7}{90}\pi^4\right)
	\nonumber &\\&
	+\frac{1}{\e}\left(-\frac{46055}{192}+\frac{7591}{576}\pi^2+\frac{473}{6}\zeta_{3}-\frac{\pi^4}{90}+\frac{27}{8}\pi^2\zeta_{3}-44 \zeta_{5}\right)
	\nonumber &\\&
	-\frac{379673}{192}+\frac{63601}{576}\pi^2+\frac{2185}{4}\zeta_{3}+\frac{133}{216}\pi^4
	\nonumber &\\&                
        -\frac{565}{24}\pi^2\zeta_{3}+\frac{1297}{8}\zeta_{5}+\frac{22229}{272160}\pi^6+\frac{345}{4}\zeta_{3}^2 + \mathcal{O}(\e)\,. &
\end{flalign}

\section{Comments on the results}
\label{sec:checks}

\subsection{Total cross section at ${\cal O}(\alpha_{s}^3)$}

The natural check for our results is the complete cancellation of all infrared singularities in the total cross section at
$\mathcal{O}(\alpha_s^3)$, that is, the sum of the layers~\eqref{eq:X23ansatzA}, \eqref{eq:X23ansatzB}, \eqref{eq:X32ansatz1A}, \eqref{eq:X32ansatz1B},
\eqref{eq:X32ansatz2},
\eqref{eq:X41ansatz1},
\eqref{eq:X41ansatz2},
\eqref{eq:X41ansatz3},
\eqref{eq:X50ansatz1},
\eqref{eq:X50ansatz2},
\eqref{eq:X50ansatz3}. We achieve the cancellation of the poles and recover the N$^3$LO coefficient of the $R$-ratio~\cite{Baikov:2012zn}:
\begin{flalign}
  R\Big|_{\alpha_s^3} &= \left(\frac{\alpha_s}{2\pi}\right)^3
  \frac{ {\cal T}^{(8)}_{q\bar{q}} + {\cal T}^{(8)}_{q\bar{q}g} + {\cal T}^{(8)}_{ggg}
    + {\cal T}^{(8)}_{q\bar{q}gg} + {\cal T}^{(8)}_{q\bar{q}q'\bar{q}'} + {\cal T}^{(8)}_{q\bar{q}q\bar{q}}
    + {\cal T}^{(8)}_{q\bar{q}ggg} + {\cal T}^{(8)}_{q\bar{q}q'\bar{q}'g} + {\cal T}^{(8)}_{q\bar{q}q\bar{q}g} }
       { {\cal T}^{(2)}_{q\bar{q}} }
  \nonumber &\\
  &= \left(\frac{\alpha_s}{2\pi}\right)^3 \left( N - \frac{1}{N} \right)
  \Bigg[
    N^2 \left( \frac{346201}{3456} - \frac{121}{144}\pi^2 - \frac{6761}{72} \zeta_3 + \frac{55}{3} \zeta_5 \right)
    \nonumber &\\& \phantom{=\left(\frac{\alpha_s}{2\pi}\right)^3\left(N-\frac{1}{N}\right)\Bigg[}
      + \frac{323}{64} + \frac{143}{8} \zeta_3 - \frac{55}{2} \zeta_5
      + \frac{1}{N^2} \left( - \frac{69}{128} \right)
  \nonumber &\\& \phantom{=\left(\frac{\alpha_s}{2\pi}\right)^3\left(N-\frac{1}{N}\right)\Bigg[}
    + \NF N \left( -\frac{62863}{1728} + \frac{11}{36}\pi^2 + \frac{1067}{36} \zeta_3 - \frac{10}{3} \zeta_5 \right)
      \nonumber &\\& \phantom{=\left(\frac{\alpha_s}{2\pi}\right)^3\left(N-\frac{1}{N}\right)\Bigg[}
        + \frac{\NF}{N} \left( \frac{29}{64} + 5\zeta_5 -\frac{19}{4}\zeta_3\right)
        + \NF^2 \left( \frac{151}{54} -\frac{1}{36}\pi^2 - \frac{19}{9}\zeta_3 \right)
      \nonumber &\\& \phantom{=\left(\frac{\alpha_s}{2\pi}\right)^3\left(N-\frac{1}{N}\right)\Bigg[}        
    + \NFZ \left(N - \frac{4}{N}\right) \left( \frac{11}{24} - \zeta_3 \right)
    \Bigg]\,.
\end{flalign}

\subsection{Singlet contribution}

The singlet contribution proportional to $(\sum_q e_q)^2$ appears for the first time at three-loop order. In the two-parton final state it arises from the three-loop quark form factor and is finite due to the absence of counterterms for the associated diagrams \cite{Baikov:2009bg}. Analogously, the singlet term is also finite in the three-parton final state case. Due to Furry's theorem, neither the~$\gamma^*\to q\bar{q}g$ nor the~$\gamma^* \to ggg$ sub-processes allow for lower-loop or lower-multiplicity counterterms which could accommodate infrared singularities. 

The four- and five-parton final states exhibit a $\e^{-1}$ pole in the singlet
term. The pole is due to a real or virtual infrared
gluon and cancels in the sum of the two contributions.
Such single pole proportional to the quartic Casimir is present also in the
one-loop triple collinear splitting function~\cite{Catani:2003vu}.
The real emission counterpart of this singularity is found in the antisymmetric
tripole contribution to the soft current for a soft gluon-quark-antiquark
emission~\cite{DelDuca:2022noh,Catani:2022hkb}.
The absence of a $\e^{-2}$ pole is explained by the fact that the singlet part
of the tree-level matrix element squared for $\gamma^* \to q\bar{q}q'\bar{q}'$
vanishes~\cite{Gehrmann-DeRidder:2004ttg}, ${\cal
  T}^{(6)}_{q\bar{q}q'\bar{q}'}\Big|_{\NFZ}=0$, as also noted
in~\cite{Catani:2003vu}.

\section{Conclusions and outlook}
\label{sec:conclusions}

In this paper, we presented analytic results for the integration over the 
respective inclusive phase space of all the contributions to $e^{+} e^{-}$ annihilation to hadrons at order 
${\cal O}(\alpha_s^3)$. A common strategy was used for all layers of the calculation, 
exploiting reverse unitarity relations to compute phase space integrals. The cancellation of infrared 
singularities and the recovery of the known finite result at N$^3$LO provides strong checks 
on our calculations. We provide expressions for the results in Section \ref{sec:results} and in Appendix \ref{app:A} in \code{FORM} format in the ancillary files. 

It will be interesting to analyze the infrared structure of our
  result in more detail. The poles of the purely virtual ${\cal O}(\alpha_s^3)$ two-parton
  corrections can be predicted by means of universal infrared factorization
  formulae~\cite{Becher:2009cu,Gardi:2009qi}. Some infrared singular terms in the
  three-, four- and five-parton final states are related to individual
  terms in the infrared factorization formula. Other infrared terms can be
  obtained from the integration of known
  ingredients~\cite{Badger:2004uk,Duhr:2014nda,Duhr:2013msa,Li:2013lsa,Dixon:2019lnw,Czakon:2022dwk,Catani:2003vu,Catani:2021kcy,Zhu:2020ftr,DelDuca:2019ggv,DelDuca:2020vst,Catani:2019nqv,DelDuca:2022noh,Catani:2022hkb,Czakon:2022fqi},
  and they should cancel between the three-, four- and five-parton final
  states. We leave this investigation to future work.

Our results represent the first step towards a future extension of the antenna subtraction 
method at N$^3$LO. In particular, from the analytic expressions obtained here, it is possible to directly 
read off the integrated form of N$^3$LO quark-antiquark antenna functions, in the configuration where 
the hard quark-antiquark pair is emitted in the final state. We envisage future work for the
calculation of analogous phase space integrals in the gluon-gluon and quark-gluon case.

\section*{Acknowledgements}

We are grateful to Thomas Gehrmann for his input, feedback and encouragement to pursue this work. We thank Xuan Chen, Aude Gehrmann-De Ridder, Nigel Glover and \mbox{Tong-Zhi Yang} for discussions and comments on the manuscript. This work was supported by the Swiss National Science Foundation (SNF) under contract 200020-204200 and by the European Research Council (ERC) under the European Union's Horizon 2020 research and innovation programme grant agreement 101019620 (ERC Advanced Grant TOPUP).

\begin{appendices}

  \section{Renormalisation of amplitudes}\label{app:ren}

  Expanding $\alpha_s$ according to~\eqref{eq:alfaren}, the renormalised amplitudes are
  \begin{equation}
    |{\cal M}^{(1)}\rangle_{q\bar{q}} = |{\cal M}^{(1),U}\rangle_{q\bar{q}}\,,
  \end{equation}
  \begin{equation}
    |{\cal M}^{(2)}\rangle_{q\bar{q}} = |{\cal M}^{(2),U}\rangle_{q\bar{q}} - \frac{\beta_0}{\e} |{\cal M}^{(1),U}\rangle_{q\bar{q}}\,,
  \end{equation}
  \begin{equation}
    |{\cal M}^{(3)}\rangle_{q\bar{q}} = |{\cal M}^{(3),U}\rangle_{q\bar{q}} - \frac{2\beta_0}{\e} |{\cal M}^{(2),U}\rangle_{q\bar{q}}
    + \left(\frac{\beta_0^2}{\e^2} - \frac{\beta_1}{2\e} \right) |{\cal M}^{(1),U}\rangle_{q\bar{q}}\,,
  \end{equation}
  \begin{equation}
    |{\cal M}^{(1)}\rangle_{q\bar{q}g} = |{\cal M}^{(1),U}\rangle_{q\bar{q}g} - \frac{\beta_0}{2\e} |{\cal M}^{(0)}\rangle_{q\bar{q}g}\,,
  \end{equation}
  \begin{equation}
    |{\cal M}^{(2)}\rangle_{q\bar{q}g} = |{\cal M}^{(2),U}\rangle_{q\bar{q}g} - \frac{3\beta_0}{2\e} |{\cal M}^{(1),U}\rangle_{q\bar{q}g}
    + \left(\frac{3}{8}\frac{\beta_0^2}{\e^2} - \frac{\beta_1}{4\e} \right) |{\cal M}^{(0)}\rangle_{q\bar{q}g}\,,
  \end{equation}
  \begin{equation}
    |{\cal M}^{(1)}\rangle_{q\bar{q}ij} = |{\cal M}^{(1),U}\rangle_{q\bar{q}ij} - \frac{\beta_0}{\e} |{\cal M}^{(0)}\rangle_{q\bar{q}ij}\,,
  \end{equation} 
  where the superscript $U$ denotes unrenormalised quantities.

  \section{NNLO results}\label{app:A}
  
  Here we summarize the analytic results up to order ${\cal O}(\alpha_{s}^2)$~\cite{Gehrmann-DeRidder:2004ttg}, extended to weight 6.
  For the two-particle final state:
  \begin{flalign}
    \label{eq:T2}
    {\cal T}^{(2)}_{q\bar q} 
    &= \int \d \Phi_2 \,\langle{\cal M}^{(0)}|{\cal M}^{(0)}\rangle_{q\bar q}  
    = 4 N (1-\e) q^2 P_2 \,, &
  \end{flalign}
  \begin{flalign}
    \label{eq:T4}
    {\cal T}^{(4)}_{q\bar q}  =& 
    \int \d \Phi_2 \,
    2\,\Re\big[\langle{\cal M}^{(1)}|{\cal M}^{(0)}\rangle_{q\bar q}\big] \nonumber &\\
    =& \left( N-\frac{1}{N}\right){\cal T}^{(2)}_{q\bar q} 
    \Bigg[ -\frac{1}{\e^2} - \frac{3}{2\e} - 4 + 
    \frac{7\pi^2}{12} 
    + \left( -8 + \frac{7\pi^2}{8} + \frac{7}{3}\zeta_3 \right)
    \e \nonumber &\\&
    + \left( - 16
      + \frac{7\pi^2}{3}
      + \frac{7}{2}\zeta_3
      - \frac{73\pi^4}{1440} \right) \e^2
    \nonumber &\\
    &
    + \e^3 \Bigg(
    - 32
    + \frac{14\pi^2}{3}
    + \frac{28\zeta_3}{3}        
    - \frac{73\pi^4}{960}
    - \frac{49\zeta_3\pi^2}{36}
    + \frac{31\zeta_5}{5}
    \Bigg)         
    \nonumber &\\	&
    + \e^4 \Bigg(
    - 64
    + \frac{28\pi^2}{3}
    + \frac{56\zeta_3}{3}          
    - \frac{73\pi^4}{360}
    - \frac{49\zeta_3\pi^2}{24}
    \nonumber &\\& \phantom{+\frac{1}{\e}\bigg(\;\;\;}
    + \frac{93\zeta_5}{10}                    
    - \frac{437\pi^6}{120960}
    - \frac{49\zeta_3^2}{18}
    \Bigg)	
    + {\cal O}(\e^5) \Bigg] \,,&
  \end{flalign}
  \begin{flalign}
    {\cal T}_{q\bar q}^{(6,[2\times 0])}  =& 
    \int \d \Phi_2 \,
    2\,\Re\big[\langle{\cal M}^{(2)}|{\cal M}^{(0)}\rangle_{q\bar q}\big]
    \nonumber &\\ = &
    \left( N-\frac{1}{N}\right)\;{\cal T}^{(2)}_{q\bar q} \;\Bigg\{
    N\, {\cal T}_{q\bar q}^{(6,[2\times 0])}\Big|_{N}
    + \frac{1}{N}\,{\cal T}_{q\bar q}^{(6,[2\times 0])}\Big|_{1/N}
    \nonumber &\\& \phantom{\left( N-\frac{1}{N}\right)\;{\cal T}^{(2)}_{q\bar q} \;\Bigg\{}
    + \NF\,\,{\cal T}_{q\bar q}^{(6,[2 \times 0])}\Big|_{\NF}
    \Bigg\}\,,
  \end{flalign}
  with
  \begin{flalign}
    {\cal T}_{q\bar q}^{(6,[2\times 0])}\Big|_{N} =& 
    \frac{1}{4\e^4} + \frac{17}{8\e^3}
    + \frac{1}{\e^2} \left( \frac{433}{144} -\frac{\pi^2}{2} \right)
    \nonumber &\\
    &+ \frac{1}{\e} \left( \frac{4045}{864} -\frac{83\pi^2}{48} 
      +\frac{7}{12}\zeta_3 \right)
    +\left(-\frac{9083}{5184}
      - \frac{2153\pi^2}{864}
      + \frac{13}{9}\zeta_3           + \frac{263\pi^4}{1440}
    \right) 
    \nonumber &\\&+\eps\Bigg(
    - \frac{1244339}{31104}
    - \frac{1943\pi^2}{5184}
    + \frac{4235\zeta_3}{216}        
    + \frac{389\pi^4}{720}
    - \frac{13\zeta_3\pi^2}{8}
    + \frac{163\zeta_5}{20}        
    \Bigg)
    \nonumber &\\
    &	+\e^2\Bigg(	
    - \frac{36528395}{186624}
    + \frac{611833\pi^2}{31104}
    + \frac{109019\zeta_3}{1296}        
    + \frac{38519\pi^4}{25920}
    - \frac{2087\zeta_3\pi^2}{216}        
    \nonumber &\\& \phantom{+\frac{1}{\e}\bigg(\;\;\;}
    + \frac{529\zeta_5}{15}
    - \frac{631\pi^6}{15120}
    - \frac{403\zeta_3^2}{36}\Bigg) + {\cal O}(\e^3)\,,
  \end{flalign}
  \begin{flalign}
    {\cal T}_{q\bar q}^{(6,[2\times 0])}\Big|_{1/N} =& 
    -\frac{1}{4\e^4} - \frac{3}{4\e^3}
    + \frac{1}{\e^2} \left( -\frac{41}{16} +\frac{13\pi^2}{24} \right)
    \nonumber &\\
    &+ \frac{1}{\e} \left( -\frac{221}{32} +\frac{3\pi^2}{2} 
      +\frac{8}{3}\zeta_3 \right)
    +\left(-\frac{1151}{64}
      + \frac{475\pi^2}{96}
      + \frac{29}{4}\zeta_3   - \frac{59\pi^4}{288}
    \right) 
    \nonumber &\\&+\eps\Bigg(
    - \frac{5741}{128}
    +\frac{ 813\pi^2}{64}
    + \frac{839\zeta_3}{24}        
    - \frac{61\pi^4}{160}
    + \frac{23\zeta_5}{5}
    - \frac{55\zeta_3\pi^2}{9}
    \Bigg)
    \nonumber &\\
    &	+\e^2\Bigg(	
    - \frac{27911}{256}
    + \frac{3991\pi^2}{128}
    + \frac{6989\zeta_3}{48}        
    - \frac{4399\pi^4}{5760}
    - \frac{125\zeta_3\pi^2}{8}
    \nonumber &\\& \phantom{+\frac{1}{\e}\bigg(\;\;\;}
    + \frac{231\zeta_5}{20}        
    - \frac{571\pi^6}{8640}
    - \frac{326\zeta_3^2}{9}\Bigg) + {\cal O}(\e^3)\,,
  \end{flalign}
  \begin{flalign}
    {\cal T}_{q\bar q}^{(6,[2\times 0])}\Big|_{\NF} =& 
    -\frac{1}{4\e^3} - \frac{1}{9\e^2}
    + \frac{1}{\e} \left( \frac{65}{216} +\frac{\pi^2}{24} \right)
    + \left( \frac{4085}{1296} -\frac{91\pi^2}{216} 
      +\frac{1}{18}\zeta_3 \right)   
    \nonumber &\\&+\eps\Bigg(
    + \frac{108653}{7776}
    - \frac{2875\pi^2}{1296}
    - \frac{119\zeta_3}{54}         
    + \frac{\pi^4}{720}
    \Bigg)
    \nonumber &\\
    &	+\e^2\Bigg(
    + \frac{2379989}{46656}
    - \frac{70855\pi^2}{7776}
    - \frac{3581\zeta_3}{324}         
    + \frac{311\pi^4}{5184}
    \nonumber &\\& \phantom{+\frac{1}{\e}\bigg(\;\;\;}
    + \frac{47\zeta_3\pi^2}{54}
    - \frac{59\zeta_5}{30}         
    \Bigg) + {\cal O}(\e^3)\,,
  \end{flalign}
  and
  \begin{flalign}
    {\cal T}_{q\bar q}^{(6,[1\times 1])}  =& 
    \int \d \Phi_2 \, \langle{\cal M}^{(1)}|{\cal M}^{(1)}\rangle_{q\bar q}  \nonumber &\\
    =
    &\,{\cal T}^{(2)}_{q\bar q}\;\left( N-\frac{1}{N}\right)^2\;
    \Bigg[ \frac{1}{4\e^4} + \frac{3}{4\e^3}
    + \frac{1}{\e^2} \left( \frac{41}{16} -\frac{\pi^2}{24} \right)
    \nonumber &\\
    &+ \frac{1}{\e} \left( 7 -\frac{\pi^2}{8} 
      -\frac{7}{6}\zeta_3 \right)
    +\left(18
      - \frac{41\pi^2}{96}
      - \frac{7}{2}\zeta_3           - \frac{7\pi^4}{480}
    \right) 
    \nonumber &\\&+\eps\Bigg(
    + 44
    - \frac{7\pi^2}{6}
    - \frac{287\zeta_3}{24}          
    - \frac{7\pi^4}{160}
    + \frac{7\zeta_3\pi^2}{36}          
    - \frac{31\zeta_5}{10}
    \Bigg)
    \nonumber &\\
    &	+\e^2\Bigg(
    + 104
    - 3\pi^2
    - \frac{98\zeta_3}{3}          
    - \frac{287\pi^4}{1920}
    + \frac{7\zeta_3\pi^2}{12}          
    \nonumber &\\& \phantom{+\frac{1}{\e}\bigg(\;\;\;}
    - \frac{93\zeta_5}{10}
    - \frac{31\pi^6}{12096}          
    + \frac{49\zeta_3^2}{18}
    \Bigg)
    + {\cal O}(\e)  \Bigg].
  \end{flalign}
  We notice a typo in~(4.10) of~\cite{Gehrmann-DeRidder:2004ttg}: the $1/\e$
  coefficient in the previous equation features a $-7/6\,\zeta_3$ instead of
  $+7/6\,\zeta_3$.

  \noindent
  For the three-particle final state:
  \begin{flalign}
    {\cal T}^{(4)}_{q\bar q g}  =&\,8\pi^2 \int \d \Phi_3 
    \langle{\cal M}^{(0)}|{\cal M}^{(0)}\rangle_{q\bar q g} \nonumber &\\
    = &
    \left( N-\frac{1}{N}\right)\;{\cal T}^{(2)}_{q\bar q} \;
    \Bigg[
    \frac{1}{\e^2} + \frac{3}{2\e} + \frac{19}{4} - 
    \frac{7\pi^2}{12} 
    + \e \left( \frac{109}{8} - \frac{7\pi^2}{8} - \frac{25}{3}\zeta_3 \right)
    \nonumber &\\ 
    &+ \e^2 \left( \frac{639}{16}
      - \frac{133\pi^2}{48}
      - \frac{25}{2}\zeta_3
      - \frac{71\pi^4}{1440} \right) 
    \nonumber &\\ 
    &+ \e^3 \left(
      \frac{3789}{32} - \frac{763\pi^2}{96} - \frac{475 \zeta_3}{12} - \frac{71\pi^4}{960}
      + \frac{175 \pi^2\zeta_3}{36} - \frac{241\zeta_5}{5}
    \right)
    \nonumber &\\ 
    &+ \e^4 \Bigg(
    \frac{22599}{64}
    - \frac{1491\pi^2}{64}
    - \frac{2725\zeta_3}{24}        
    - \frac{1349\pi^4}{5760}
    + \frac{175\pi^2\zeta_3}{24}
    \nonumber &\\ & \phantom{+ \e^4 \Bigg(}
    - \frac{723\zeta_5}{10}
    - \frac{4027\pi^6}{120960}
    + \frac{625\zeta_3^2}{18}
    \Bigg)
    + {\cal O}(\e^3) \Bigg] \,,
  \end{flalign} 
  \begin{flalign}
    {\cal T}^{(6)}_{q\bar q g}  =&\, 8\pi^2 \int \d \Phi_32\,\Re\big[\langle{\cal M}^{(1)}|{\cal M}^{(0)}\rangle_{q\bar qg}\big]
    \nonumber &\\ =&
    \left( N-\frac{1}{N}\right)\;{\cal T}^{(2)}_{q\bar q} \;
    \Bigg\{
    N\, {\cal T}_{q\bar qg}^{6}\Big|_{N}
    + \frac{1}{N}\,{\cal T}_{q\bar qg}^{6}\Big|_{1/N}
    + \NF\,\,{\cal T}_{q\bar qg}^{6}\Big|_{\NF}
    \Bigg\},
  \end{flalign}
  with
  \begin{flalign}
    {\cal T}^{(6)}_{q\bar q g}\Big|_{N}  =&
    -\frac{5}{4\e^4} - \frac{67}{12\e^3}
    + \frac{1}{\e^2} \left( -\frac{141}{8} +\frac{13\pi^2}{8} \right)
    \nonumber &\\
    &+ \frac{1}{\e} \left( -\frac{1481}{24} +\frac{107\pi^2}{18} 
      +\frac{55}{3}\zeta_3 \right)
    +\left(-\frac{10385}{48}
      + \frac{64\pi^2}{3}
      + \frac{1265}{18}\zeta_3           - \frac{41\pi^4}{96}
    \right) 
    \nonumber &\\&
    +\e\left(
      -\frac{24915}{32}
      + \frac{22223\pi^2}{288}
      + \frac{1517\zeta_3}{6}        
      - \frac{10289\pi^4}{8640}
      - \frac{149\pi^2\zeta_3}{6}
      + 143\zeta_5
    \right)
    \nonumber &\\&
    +\e^2\Bigg(
    -\frac{183957}{64}
    + \frac{162077\pi^2}{576}
    + \frac{68069\zeta_3}{72}        
    - \frac{28493\pi^4}{5760}
    - \frac{18017\pi^2\zeta_3}{216}
    \nonumber &\\& \phantom{+\frac{1}{\e}\bigg(\;\;\;}
    + \frac{15521\zeta_5}{30}
    + \frac{5357\pi^6}{60480}        
    - \frac{1345\zeta_{3}^{2}}{9}
    \Bigg)  + {\cal O}(\e^3)\,, &
  \end{flalign}
  \begin{flalign}
    {\cal T}^{(6)}_{q\bar q g}\Big|_{1/N}  =&
    \frac{1}{\e^4} + \frac{3}{\e^3}
    + \frac{1}{\e^2} \left( \frac{93}{8} -\frac{4\pi^2}{3} \right)
    \nonumber &\\
    &+ \frac{1}{\e} \left( \frac{79}{2} -\frac{15\pi^2}{4} 
      -\frac{53}{3}\zeta_3 \right)
    +\left(\frac{1069}{8}
      - \frac{697\pi^2}{48}
      - \frac{91}{2}\zeta_3   + \frac{19\pi^4}{72}
    \right) 
    \nonumber &\\&
    +\e\left(
      \frac{3653}{8}
      - 50\pi^2
      - \frac{723\zeta_3}{4}        
      + \frac{19\pi^4}{24}
      + \frac{493\pi^2\zeta_3}{18}
      - \frac{897\zeta_5}{5}
    \right)
    \nonumber &\\&
    +\e^2\Bigg(
    \frac{6375}{4}
    - \frac{8297\pi^2}{48}
    - \frac{3911\zeta_3}{6}        
    + \frac{8303\pi^4}{2880}
    + \frac{273\pi^2\zeta_3}{4}        
    \nonumber &\\& \phantom{+\frac{1}{\e}\bigg(\;\;\;}
    - \frac{4257\zeta_5}{10}
    - \frac{827\pi^6}{11340}
    + \frac{1931\zeta_{3}^{2}}{9}        
    \Bigg) + {\cal O}(\e^3)\,, &
  \end{flalign}
  \begin{flalign}
    {\cal T}^{(6)}_{q\bar q g}\Big|_{\NF} =&
    \frac{1}{3\e^3} + \frac{1}{2\e^2}
    + \frac{1}{\e} \left( \frac{19}{12} -\frac{7\pi^2}{36} \right)
    + \left( \frac{109}{24} -\frac{7\pi^2}{24} 
      -\frac{25}{9}\zeta_3 \right) 	
    \nonumber &\\&
    +\e\left(
      \frac{213}{16}
      - \frac{133\pi^2}{144}
      - \frac{25\zeta_3}{6}
      - \frac{71\pi^4}{4320} 
    \right)
    \nonumber &\\&
    +\e^2\left(
      \frac{1263}{32}
      - \frac{763\pi^2}{288}
      - \frac{475\zeta_3}{36}         
      - \frac{71\pi^4}{2880}
      + \frac{175\pi^2\zeta_3}{108}
      - \frac{241\zeta_5}{15}	 
    \right) + {\cal O}(\e^3)\,. &
  \end{flalign}
  We notice a typo in~(4.34) of~\cite{Gehrmann-DeRidder:2004ttg}: the $1/\e$
  coefficient in the previous equation features a $19/12$ instead of $19/2$.
  This typo was fixed in~\cite{Gehrmann-DeRidder:2005btv}.
  
  \noindent
  For the four-particle final state:
  \begin{flalign}
    {\cal T}^{(6)}_{q\bar q gg}  =&
    \,64 \pi^4 \,
    \int \d \Phi_4  \langle{\cal M}^{(0)}|{\cal M}^{(0)}\rangle_{q\bar q gg} \nonumber \\
    =&\,
    \left( N-\frac{1}{N}\right)\;{\cal T}^{(2)}_{q\bar q} \;
    \Bigg\{ N\,{\cal T}^{(6)}_{q\bar q gg}\Big|_{N} + \frac{1}{N}\,{\cal T}^{(6)}_{q\bar q gg}\Big|_{1/N} \Bigg\},
  \end{flalign}
  with
  \begin{flalign}
    {\cal T}^{(6)}_{q\bar q gg}\Big|_{N}  =&
    \frac{3}{4\e^4} + \frac{65}{24\e^3}
    + \frac{1}{\e^2} \left( \frac{217}{18} -\frac{13\pi^2}{12} \right)
    \nonumber &\\&
    + \frac{1}{\e} \left( \frac{43223}{864} -\frac{589\pi^2}{144} 
      -\frac{71}{4}\zeta_3 \right)
    +\left(\frac{1076717}{5184}
      - \frac{7955\pi^2}{432}
      - \frac{1327}{18}\zeta_3   + \frac{373\pi^4}{1440}
    \right) 
    \nonumber &\\
    &
    +\e\left(
      \frac{26964431}{31104}
      - \frac{398557\pi^2}{5184}
      - \frac{73301\zeta_3}{216}        
      + \frac{5207\pi^4}{8640}
      + \frac{1891\pi^2\zeta_3}{72}
      - \frac{2661\zeta_5}{20}
    \right)
    \nonumber &\\
    &
    +\e^2\Bigg(
    \frac{677415461}{186624}
    - \frac{9919003\pi^2}{31104}
    - \frac{1857845\zeta_3}{1296}        
    + \frac{2975\pi^4}{1296}
    \nonumber &\\& \phantom{+\frac{1}{\e}\bigg(\;\;\;}
    + \frac{11969\pi^2\zeta_3}{108}                
    - \frac{4211\zeta_5}{6}
    - \frac{139\pi^6}{30240}
    + \frac{2723\zeta_3^2}{12}        
    \Bigg) 	+ {\cal O}(\e^3) \,,
  \end{flalign}
  \begin{flalign}
    {\cal T}^{(6)}_{q\bar q gg}\Big|_{1/N}  =&
    -\frac{1}{2\e^4} - \frac{3}{2\e^3}
    + \frac{1}{\e^2} \left( -\frac{13}{2} +\frac{3\pi^2}{4} \right)
    + \frac{1}{\e} \left( -\frac{845}{32} +\frac{9\pi^2}{4} 
      +\frac{40}{3}\zeta_3 \right) \nonumber &\\&
    +\left(-\frac{6921}{64}
      + \frac{473\pi^2}{48}
      + 40\zeta_3           - \frac{17\pi^4}{144}
    \right) 
    \nonumber &\\
    &
    +\e\left(
      -\frac{57305}{128}       
      + \frac{7735\pi^2}{192}
      + \frac{4295\zeta_3}{24}        
      - \frac{17\pi^4}{48}
      - \frac{121\pi^2\zeta_3}{6}
      + \frac{636\zeta_5}{5}
    \right)
    \nonumber &\\
    &
    +\e^2\Bigg(
    -\frac{477601}{256}
    + \frac{21125\pi^2}{128}
    + \frac{35555\zeta_3}{48}        
    - \frac{107\pi^4}{80}
    - \frac{121\pi^2\zeta_3}{2}        
    \nonumber &\\& \phantom{+\frac{1}{\e}\bigg(\;\;\;}
    + \frac{1908\zeta_5}{5}
    + \frac{4763\pi^6}{90720}
    - \frac{3281\zeta_3^2}{18}        
    \Bigg)	+ {\cal O}(\e^3) \,,
  \end{flalign}
  \begin{flalign}
    {\cal T}^{(6)}_{q\bar q q' \bar q'}  =&\,
    64 \pi^4 \,\int \d \Phi_4  \langle{\cal M}^{(0)}|
    {\cal M}^{(0)}\rangle_{q\bar q q'\bar q'} \nonumber &\\
    = &\,
    {\cal T}^{(2)}_{q\bar q} \;\left( N-\frac{1}{N}\right)\; 
    (\NF-1)\,
    \Bigg[ 
    - \frac{1}{12\e^3} - \frac{7}{18\e^2}
    + \frac{1}{\e} \left( -\frac{407}{216} +\frac{11\pi^2}{72} \right) \nonumber &\\&
    + \left( - \frac{11753}{1296} +\frac{77\pi^2}{108} 
      +\frac{67}{18}\zeta_3 \right) 	
    +\e\left(
      -\frac{340475}{7776}
      + \frac{4369\pi^2}{1296}
      + \frac{469\zeta_3}{27}        
      + \frac{137\pi^4}{4320}
    \right)
    \nonumber &\\
    &
    +\e^2\Bigg(
    -\frac{9739325}{46656}
    + \frac{120859\pi^2}{7776}
    + \frac{25811\zeta_3}{324}        
    + \frac{959\pi^4}{6480}
    \nonumber &\\& \phantom{+\frac{1}{\e}\bigg(\;\;\;}
    - \frac{629\pi^2\zeta_3}{108}        
    + \frac{1651\zeta_5}{30}
    \Bigg)		 
    + {\cal O}(\e^3) \Bigg], &
  \end{flalign}
  and
  \begin{flalign}
    {\cal T}^{(6)}_{q\bar q q \bar q}  =&\,
    64 \pi^4 \,
    \int \d \Phi_4  \langle{\cal M}^{(0)}|{\cal M}^{(0)}\rangle_{q\bar q q\bar q} \nonumber &\\
    =& \frac{1}{\NF-1} {\cal T}^{(6)}_{q\bar q q' \bar q'} 
    + \left( N-\frac{1}{N}\right) \;{\cal T}^{(2)}_{q\bar q}\;  \frac{1}{N} \,\Bigg[ 
    \frac{1}{\e} \left( \frac{13}{16} -\frac{\pi^2}{8} + 
      \frac{1}{2}\zeta_3\right) \nonumber &\\
    &
    + \left(  \frac{339}{32} -\frac{17\pi^2}{24} 
      -\frac{21}{4}\zeta_3 +\frac{2\pi^4}{45} \right)  
    \nonumber &\\
    &
    +\e\left(
      \frac{5391}{64}
      - \frac{133\pi^2}{32}
      - \frac{125\zeta_3}{4}        
      - \frac{\pi^4}{10}
      - \frac{11\pi^2\zeta_3}{12}
      +22\zeta_5
    \right)
    \nonumber &\\
    &
    +\e^2\Bigg(
    \frac{68123}{128}
    - \frac{5089\pi^2}{192}
    - \frac{3649\zeta_3}{24}        
    - \frac{7\pi^4}{10}
    + \frac{203\pi^2\zeta_3}{24}        
    \nonumber &\\& \phantom{+\frac{1}{\e}\bigg(\;\;\;}
    - \frac{357\zeta_5}{4}
    + \frac{277\pi^6}{11340}
    - \frac{95\zeta_3^2}{6}        
    \Bigg)
    + {\cal O}(\e^3) 
    \Bigg]. &
  \end{flalign}

\end{appendices}

\newpage

\bibliographystyle{JHEP}
\bibliography{main}

\end{document}